\documentclass[aps,pre,twocolumn,superscriptaddress,showpacs,amsmath,amssymb,footinbib,longbibliography]{revtex4-2}
\usepackage[english]{babel}
\usepackage{amssymb,amsbsy,amsmath}
\usepackage{graphicx}% Include figure files
\usepackage{dcolumn}% Align table columns on decimal point
\usepackage{bm}% bold math
\usepackage{subfigure}
\usepackage{here}
\usepackage{color}
\usepackage{textcomp}
\usepackage{braket}%Bra-Ket-Schreibweise
\usepackage[colorlinks,bookmarks=false,citecolor=blue,linkcolor=red,urlcolor=blue]{hyperref}
\newcommand{\op}[1]{%
    \fontdimen12\textfont3=2pt\fontdimen12\scriptfont3=1.4pt%
    \!\null\mathop{\vphantom{#1}\smash{#1}}\limits_{\sim}\null\!}
\newcommand{\xref}[1]{\protect\ref{#1}}
\newcommand{\figref}[1]{Fig.~\protect\ref{#1}}
\newcommand{\fmref}[1]{(\protect\ref{#1})}
\def\bra#1{\langle \, {#1} \, | \,}
\def\ket#1{\, | \, {#1} \, \rangle}
\renewcommand{\eqref}[1]{Eq.~(\protect\ref{#1})}

\begin{document}
\title{Toroidal magnetic molecules stripped to their basics}

\author{Daniel Pister}
\author{Kilian Irl\"ander}
\email{ORCID: 0000-0002-0223-6506}
\author{Dennis Westerbeck}
\email{ORCID: 0000-0002-8830-6551}
\author{J\"urgen Schnack}
\email{ORCID: 0000-0003-0702-2723}
\affiliation{Fakult\"at f\"ur Physik, Universit\"at Bielefeld, Postfach 100131, D-33501 Bielefeld, Germany}

\date{\today}

\begin{abstract}
Molecular magnetic toroidal moments are molecule-based structures of 
quantum spins that are expected to boost
magnetic storage technology and quantum computing. 
We study selected fictitious but typical examples of single-molecule toroidal magnet
behavior, discuss the essence of the concept and clarify inappropriate
or even wrong assignments of physical properties. We provide an outlook
that discusses necessary ingredients to the concept of toroidicity.
\end{abstract}

%\pacs{75.10.Jm,75.50.Xx,75.40.Mg} 

\maketitle

%%%%%%%%%%%%%%%%%%%%%%%%%%%%%%%%%%%%%%%%%%%%%%%%%%%%%%%%%%%%%%%%%%%%%%%%
\section{Introduction}

Magnetic molecules constitute a fascinating class of magnetic materials, see e.g.\
\cite{GSV:2006,Blundell:CP07,FuW:RMP13,LvS:CSR15,Sch:CP19}, for which in particular two properties 
give rise to hope for technological applications. The first property is the appearance
of slow relaxation of the magnetization
\cite{FST:PRL96,TLB:Nature96,GOR:N17,RGC:CS18,GDC:S18,GMR:S22}
which would 
allow to use a single molecule as a bit of magnetic storage. The main obstacle in
the context of this paper is the appearance of temperature-independent 
quantum tunneling of the magnetization due to a tunneling gap (avoided level crossing)
that opens up for non-Kramers systems for instance for non-collinear arrangements 
of easy anisotropy axes \cite{IrS:PRB20,HyS:MC22}. 

The second property is slow decoherence which would
allow to use a single molecule as a bit in quantum computing schemes
\cite{ARM:PRL07,Wer:NM07,KWW:PRB14,SKD:N16,GFB:PRL17,GLH:NC19,AtS:JACS19,CSC:PRR20,CZC:APL21,PCW:npjQI21,LMU:NP21,CPC:JPCL22}. 
Here, recent efforts
focus on clock transitions, i.e., transitions where at least the first derivative 
of the transition energy with respect to an external magnetic field is zero.
Such transitions are more robust against fluctuations of the field
than others and should thus exhibit longer coherence times.

Molecular toroidal magnetic states 
\cite{THM:ACIE06,ARS:N07,SoC:PRB08,SFM:JPCM08,ULH:JACS12,WSL:CS12,XCZ:IC12,DVG:CAEJ15,GGM:ACIE18,VLS:ACIE18,CrA:PRB18,LVG:DT19,RAS:PB20,Pav:PRB20,ZZQ:M20,ABV:EJIC21,HyS:MC22}
are often advertised as a means to improve the properties 
for the use as both units of single-molecule magnetic storage
and quantum q-bits. The basic quantum states to be manipulated
in such schemes are left and right circular (chiral)
orientations of the toroidal (ground) states.
One of the reasons for the assumed improved properties
is that toroidal arrangements of spins appear
less susceptible to fluctuating magnetic fields, at least in a
mean field picture, which should shield them from magnetic
disturbances by other spins, compare discussion in
\cite{TSL:PRB12} for the related chirality in spin triangles.
This might indeed be the case but the usability of toroidal
states and structures depends on finer details of the spin
Hamiltonian as well as on the precise coupling to disturbing
sources \cite{GCB:JPCL21,VoS:PRB20,ISS:EPJB21}.

In the following, we demonstrate that quantum spin Hamiltonians 
that consist of isotropic Heisenberg interaction terms as well as of 
toroidal arrangements of single-ion easy-axis anisotropy terms may have toroidal low-lying states
but these states do not offer any new insight 
compared to systems with simple non-collinear single-ion easy-axis anisotropy terms.
The reason is that the toroidal arrangement of single-ion easy axes
can be transformed into non-toroidal arrangements without altering the 
Hamiltonian and its spectrum. 
This insight also explains that the S-shape of magnetization curves, often
taken as hallmark of toroidal systems, is not a feature that can be used to 
unquestionably identify toroidal spin systems.

Non-Kramers toroidal spin systems, i.e., systems with integer total spin,
practically unavoidably possess a tunneling gap between their lowest 
states at $B=0$. One therefore must expect non-negligible quantum tunneling
as decades of investigations of single-molecule magnets have taught us.
We present tunneling gaps for dimeric and trimeric systems. For Kramers systems
(odd number of half-integer spins) which show no gap, transition 
rates between toroidal ground states induced by small (fluctuating) fields
may impact their stability, see Ref.~\cite{HyS:MC22} for a recent discussion.

If toroidal states should provide concepts and prospects beyond what we
already know from single-molecule magnets we have to ask which terms in a
Hamiltonian would foster a new behavior that is indeed intimately 
connected to the concept of toroidal moments. Again in line with \cite{HyS:MC22}, 
we think that dipolar interactions
as well as Dzyaloshinskii-Moriya interactions or anisotropic exchange in general 
are a prerequisite 
for a magnetization dynamics where toroidicity plays a role.

The paper is organized as follows. In Section \xref{sec-2} we discuss spin Hamiltonians 
with toroidal arrangements of easy-axis single-ion anisotropies together 
with their properties. In Section \xref{sec-3} we discuss necessary prerequisites 
for the use of toroidal moments in magnetic molecules. A summary is provided
in Sec.~\xref{sec-dc}.

%%%%%%%%%%%%%%%%%%%%%%%%%%%%%%%%%%%%%%%%%%%%%%%%%%%%%%%%%%%%%%%%%%%%%%%%
\section{Spin systems with non-collinear single-ion anisotropies}
\label{sec-2}

A typical Hamiltonian employed for magnetic molecules made of
d-elements (and used as approximation for f-elements) consists
of a Heisenberg exchange term, a term collecting the single-ion
anisotropies, and a Zeeman term, i.e.
%--------------------------------------------------------
\begin{eqnarray}
\label{E-2-1}
\op{H}
&=&
-
2\;
\sum_{i<j}\;
{J}_{ij}
\op{\vec{s}}_i \cdot \op{\vec{s}}_j
+
\sum_{i}\;
\op{\vec{s}}_i \cdot 
{\mathbf D}_i
\cdot \op{\vec{s}}_i
\\
&&+
\mu_B\, \vec{B}\cdot
\sum_{i}\;
g_i
\op{\vec{s}}_i
\nonumber
\ .
\end{eqnarray}
%--------------------------------------------------------
Here operators are marked by a tilde, and ${J}_{ij}$ denotes the
exchange parameters between spins at sites $i$ 
and $j$. A negative ${J}_{ij}$ corresponds to an
antiferromagnetic interaction, a positive one to a ferromagnetic
interaction. For the sake of simplicity it is assumed that the
spectroscopic splitting is given by numbers $g_i$.

${\mathbf D}_i$ denotes the single-ion anisotropy tensor of the spin
at site $i$ which, in its eigensystem $\vec{e}_{i}^{\,1}$,
$\vec{e}_{i}^{\,2}$, $\vec{e}_{i}^{\,3}$, can be decomposed as 
%--------------------------------------------------------
\begin{eqnarray}
\label{E-2-2}
{\mathbf D}_i
&=&
D_i \vec{e}_{i}^{\,3} \otimes \vec{e}_{i}^{\,3}
+
E_i
 \left\{
\vec{e}_{i}^{\,1} \otimes \vec{e}_{i}^{\,1}
- 
\vec{e}_{i}^{\,2} \otimes \vec{e}_{i}^{\,2}
\right\}\ .
\end{eqnarray}
%--------------------------------------------------------
In the following we will assume that the $E_i$ are zero and all
$D_i<0$, i.e.\ the anisotropy is locally of pure easy axis
type \footnote{${\mathbf D}_i$ should not be confused with a
  similar symbol denoting the Dzyaloshinskii-Moriya vector. The
  later is a vector and connects two spins.}.

Such a Hamiltonian is often employed when approximately modeling, e.g.,
dysprosium-containing magnetic molecules where the Dy moments
experience very strong easy axes \cite{VSL:NC17,SoC:PRB08,HyS:MC22}. 
This corresponds to large negative $D_i$. 

\begin{figure}[ht!]
\centering
\includegraphics*[clip,width=0.55\columnwidth]{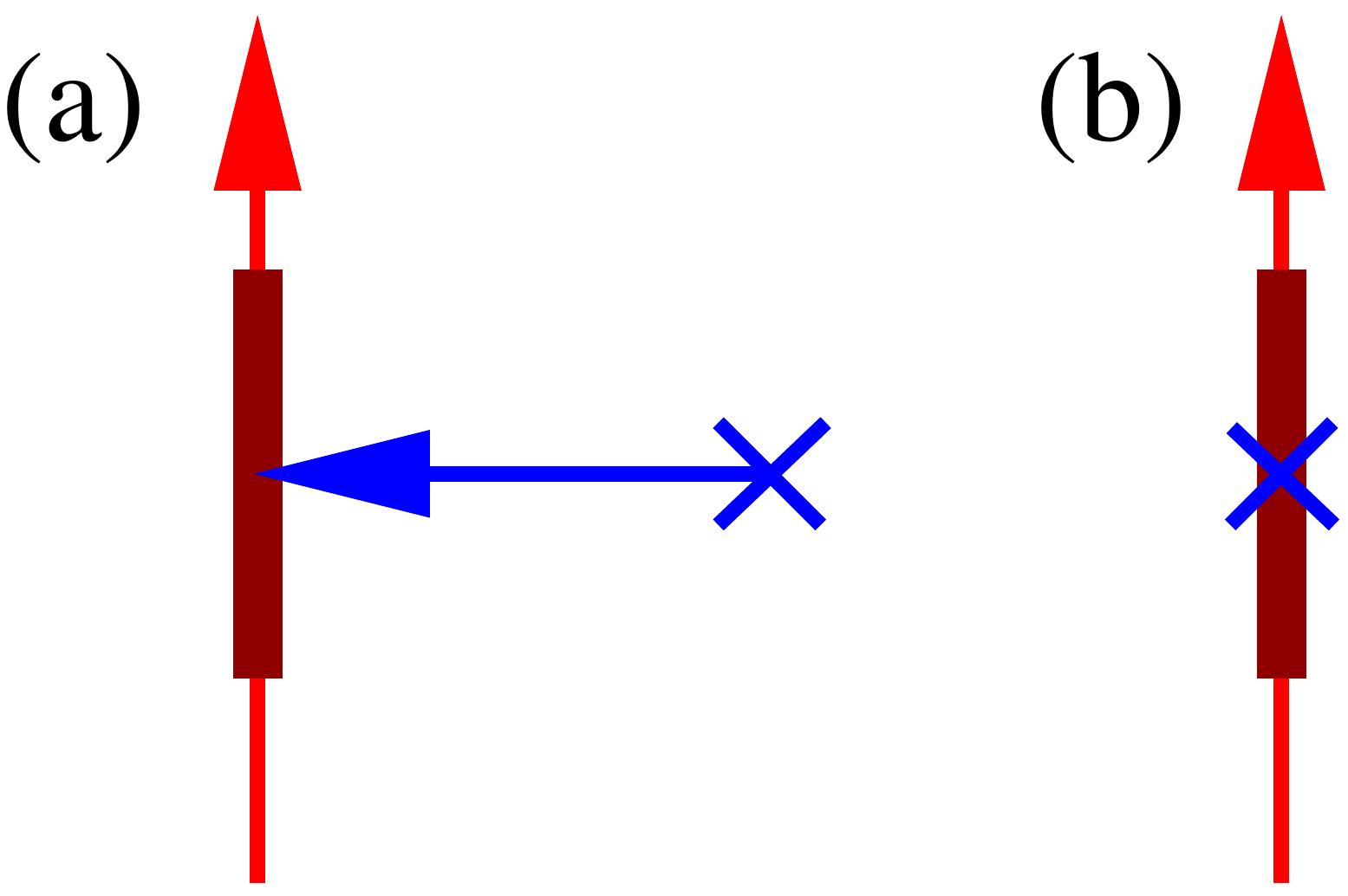}
\caption{The toroidal moment of a single spin is arbitrary due
  to the arbitrariness of the point of reference (X). In (a)
  the toroidal moment is non-zero, in (b) it is zero. The red
  arrow depicts a classical spin, the brown bar symbolizes the
  easy-axis single-ion anisotropy (for later use).} 
\label{toroidal-f-a}
\end{figure}

A toroidal moment of a set of spins is defined as
%--------------------------------------------------------
\begin{eqnarray}
\label{E-2-3}
\op{\vec{\tau}}
&=&
\sum_{i}\;
\vec{r}_i \times \op{\vec{s}}_i
\ ,
\end{eqnarray}
%--------------------------------------------------------
where the $\vec{r}_i$ are classical position vectors of the
respective spin sites with respect to a chosen point of reference.
The definition reminds one of the respective formula for the
angular momentum, and it shares with that definition the
property that the quantity contains some arbitrariness due to
the arbitrariness of the point of reference, see
\figref{toroidal-f-a} for the simple example of a single spin. 

%===================    figure   =================================
%===================    figure  =================================

%%%%%%%%%%%%%%%%%%%%%%%%%%%%%%%%%%%%%%%%%%%%%%%%%%%%%%%%%%%%%%%%%%%%%%%%
\subsection{Symmetry properties, toroidal moments, and energy spectrum}
\label{sec-2-1}

Hamiltonian \fmref{E-2-1} possesses an interesting symmetry in
view of the concept of toroidal moments. If one rotates all easy
axes as well as the field vector by the same angle about a
common axis, the energy spectrum remains the same, and so does the
magnetization as function of temperature $T$ and magnitude of
the field $B$. The reason is that the Heisenberg term is
isotropic and does not know anything about the absolute
orientation of the anisotropy tensors in space. Only the
relative orientations of the anisotropy tensors with respect to
each other matter.

This is a very important and far reaching
property since it allows to transform a toroidal moment to a
value between a minimum and a maximum by a global rotation
without changing the energy spectrum and the magnetization. In
many, in particular symmetric, cases the toroidal moment can thus
be eventually transformed to zero.

The following graphical representations show such
transformations for classical spin systems for simplicity but
symmetries and transformations extend to the respective quantum
versions. The applied field is not shown; one should keep in
mind that it has to be transformed alongside.

%===================    figure   =================================
\begin{figure}[ht!]
\centering
\includegraphics*[clip,width=0.65\columnwidth]{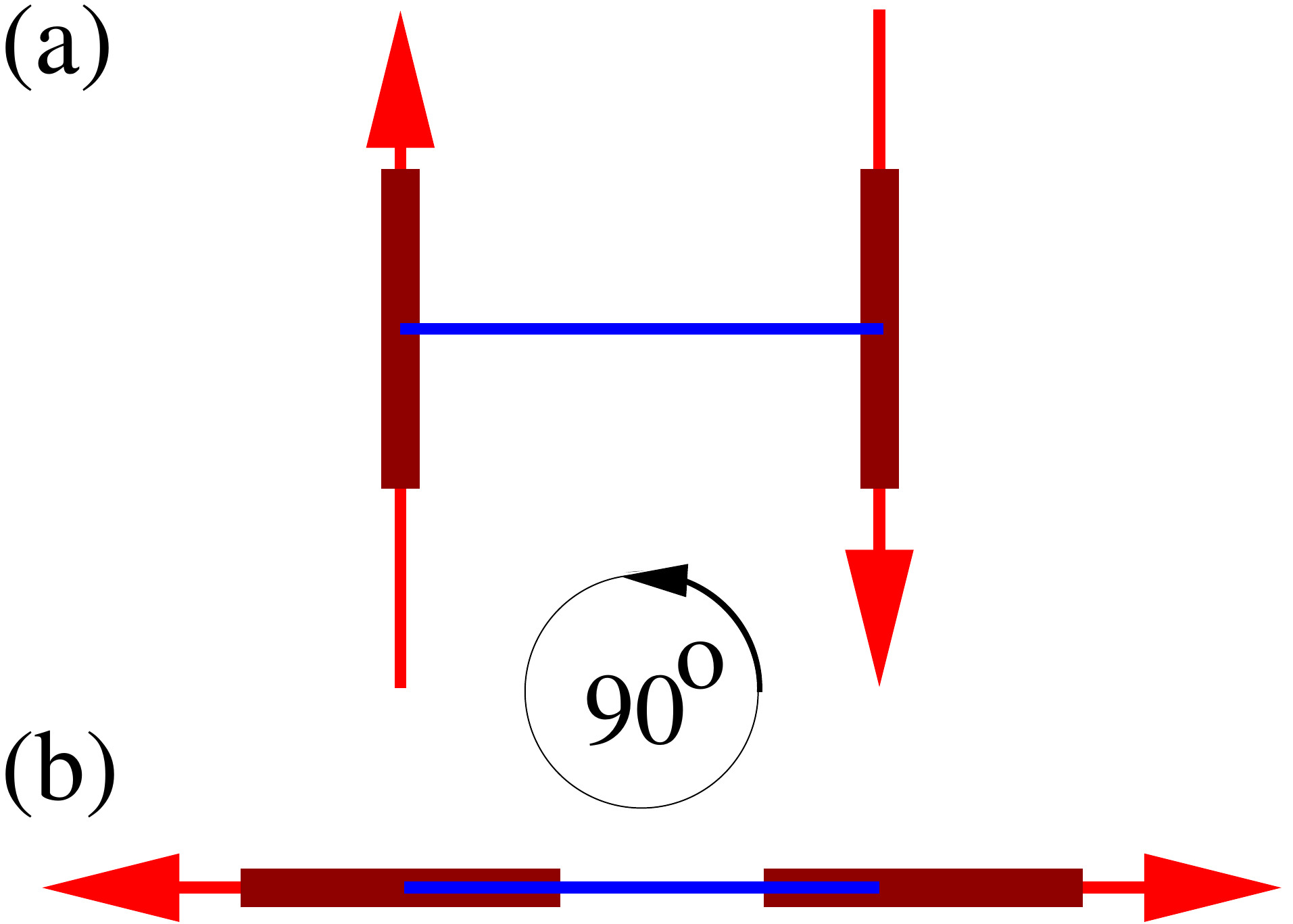}
\caption{The non-zero toroidal moment of the ground state of an
  antiferromagnetically coupled classical dimer defined with
  respect to the center between both spins (a) can be
  transformed to zero (b) by a common rotation of both easy axes
  (brown bars) by 90$^\circ$ about a common axis.} 
\label{toroidal-f-b}
\end{figure}
%===================    figure  =================================

Figure \xref{toroidal-f-b} shows in (a) the simple case of
a toroidal moment of the ground state of an
antiferromagnetically coupled classical dimer defined with
respect to the center between both spins. This moment can be
transformed to zero, compare part (b), by a common rotation of
both easy axes by 90$^\circ$ about a common axis. 

%===================    figure   =================================
\begin{figure}[ht!]
\centering
\includegraphics*[clip,width=0.55\columnwidth]{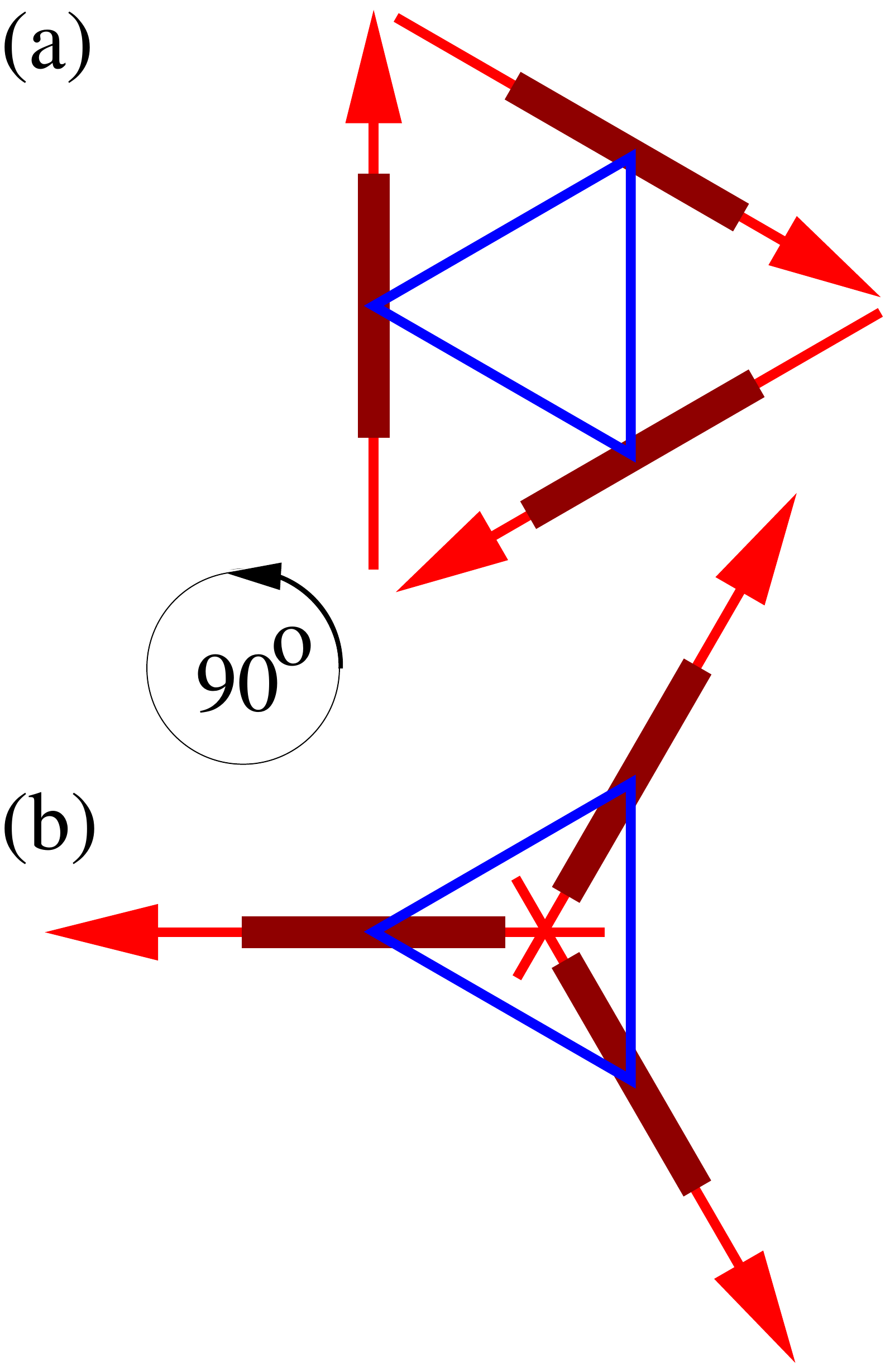}
\caption{The non-zero toroidal moment of the ground state of an
  antiferromagnetically coupled classical spin triangle defined with
  respect to the center of the triangle (a) can be
  transformed to zero (b) by a common rotation of the easy axes
  (brown bars) by 90$^\circ$ about a common axis.} 
\label{toroidal-f-c}
\end{figure}
%===================    figure  =================================

The same is true for other arrangements as for instance shown in
\figref{toroidal-f-c} for a triangular configuration that can be
transformed to zero toroidal moment without changing the energy
spectrum and the magnetization. Squares, hexagons etc.\ behave 
in the same way.

Therefore we can state that if a spin Hamiltonian contains only
Heisenberg interactions and single-ion anisotropy, the concept
of toroidicity is virtually meaningless insofar as it does not
offer any new insight into the magnetic properties of the spin
system. The energy spectrum as well as thermal expectation values 
of magnetization, susceptibility, or heat capacity remain unchanged 
under the discussed transformation, i.e., are not correlated 
at all with the expectation value of the toroidal moment.

%===================    figure   =================================
\begin{figure}[ht!]
\centering
\includegraphics*[clip,width=0.85\columnwidth]{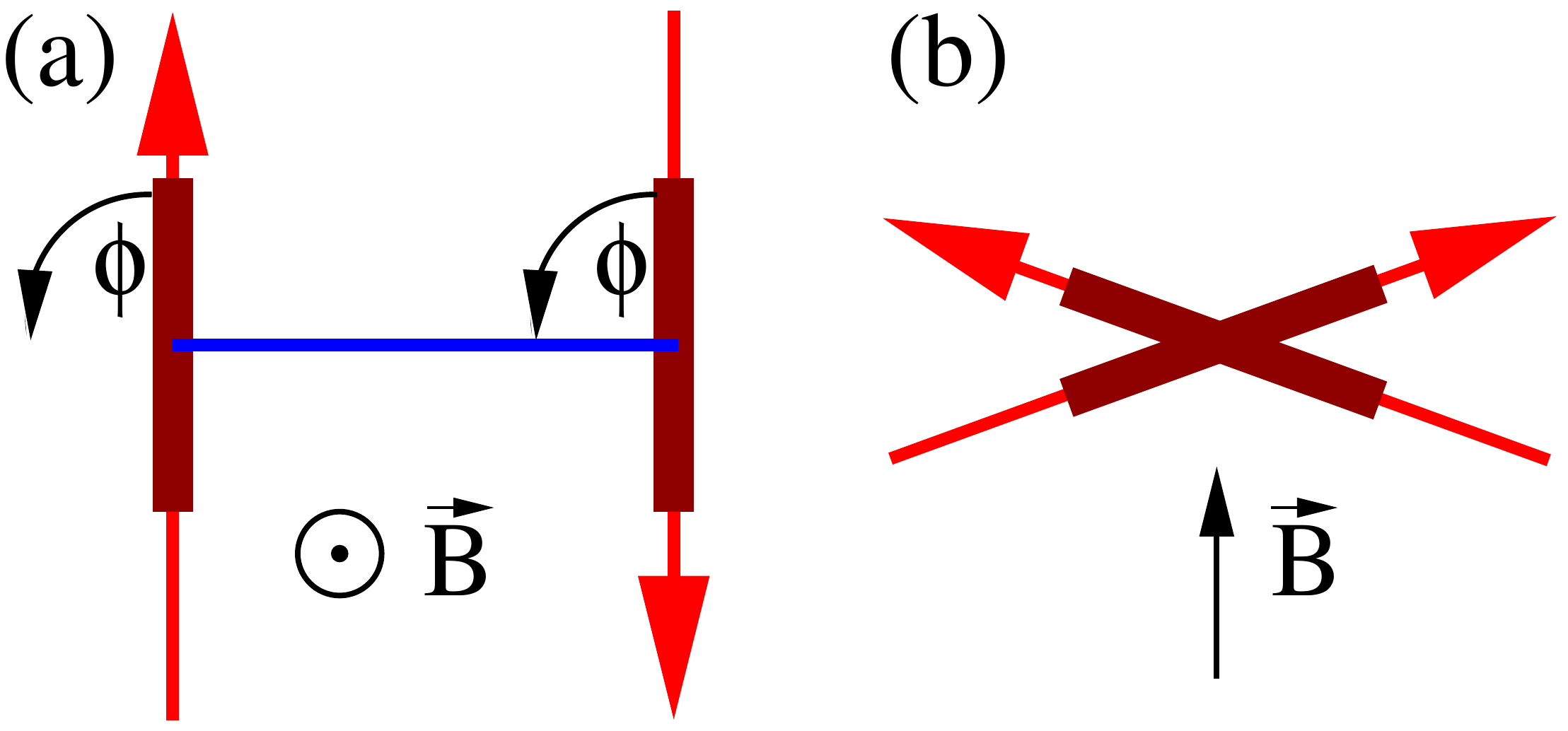}
\caption{(a) Top view of an   antiferromagnetically coupled classical dimer 
with slightly tilted easy axes
  (brown bars). The tilt angle is seen in the side view (b).
  $\phi$ denotes the angle by which the anisotropy
  axes are collectively rotated about the field axis.} 
\label{toroidal-f-d}
\end{figure}
%===================    figure  =================================

We demonstrate these statements on the simple example of a
spin dimer. The arrangement is similar to that of the
hexagonal ring in \cite{SoC:PRB08} where the easy anisotropy
axes are tilted with respect to a plane that is perpendicular to
the field along $z$-direction, compare \figref{toroidal-f-d}.

%===================    figure   =================================
\begin{figure}[ht!]
\centering
\includegraphics*[clip,width=0.85\columnwidth]{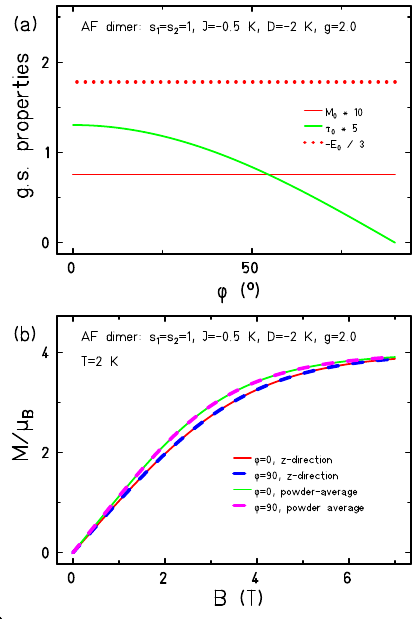}
\caption{(a) Ground state magnetization, toroidal moment, and energy for
  a magnetic field of $B=0.1$~T as a function of $\phi=0\dots
  90^\circ$ (appropriately scaled). Without loss of generality we choose typical
  parameters $J=-0.5$~K, $D_i=-2$~K, and $g_i=2.0$. The anisotropy
  axes are tilted by $10^\circ$ with respect to the plane
  perpendicular to the field axis, compare
  \figref{toroidal-f-d}(b). 
  (b) Magnetization along $z$ and powder average as function of
  field strength $B$ for $T=2$~K for the two extreme cases with $\phi=0$ and 
  $\phi=90^\circ$.} 
\label{toroidal-f-e}
\end{figure}
%===================    figure  =================================

We evaluate the toroidal moment as well as the magnetization,
which both point along $z$-direction, at $T=0$ for a small
magnetic field. This problem, by the way, can be solved 
analytically \cite{Bal:JMMM22}.
For $\phi=0$, which corresponds to the situation
shown in \figref{toroidal-f-d}, the ground state $\ket{\psi_0}$ 
has got a non-vanishing toroidal moment 
$\tau_0=\bra{\psi_0}\op{\tau}^z\ket{\psi_0}$. 
With increasing $\phi$ the toroidal moment decreases steadily
until it vanishes at $\phi=90^\circ$, see
\figref{toroidal-f-e}(a). The magnetization of the ground state,
$M_0=-g \mu_B \bra{\psi_0}\op{S}^z\ket{\psi_0}$, does not change at
all and neither does the whole energy spectrum (only ground state energy $E_0$
shown). 
This means that both the magnetization along $z$-direction as well as the
powder-averaged magnetization remain the same for all angles
$\phi$ as it must be since the Hamiltonian is not at all altered 
by the symmetry transformation. 
The result is shown in \figref{toroidal-f-e}(b), where the
magnetization curves for $\phi=0^\circ$ and $\phi=90^\circ$ are
displayed along the field.

%%%%%%%%%%%%%%%%%%%%%%%%%%%%%%%%%%%%%%%%%%%%%%%%%%%%%%%%%%%%%%%%%%%%%%%%
\subsection{Shape of magnetization curves}
\label{sec-2-2}

Sometimes, it is argued that a shape of the
low-temperature low-field magnetization curve which resembles
the letter ``S" is a signature of a toroidal moment, 
see Ref.~\cite{VSL:NC17} for an example. The same authors 
weaken their statement in Ref.~\cite{ABV:EJIC21}.
As one may deduce from the previous discussion, 
S-shaped magnetization curves cannot be taken 
as signature of toroidal moments. 
In particular, cases where the toroidal moment can be
transformed to zero without altering the magnetization curve
demonstrate unquestionably that such a simple relation cannot
exist. 

%===================    figure   =================================
\begin{figure}[ht!]
\centering
\includegraphics*[clip,width=0.85\columnwidth]{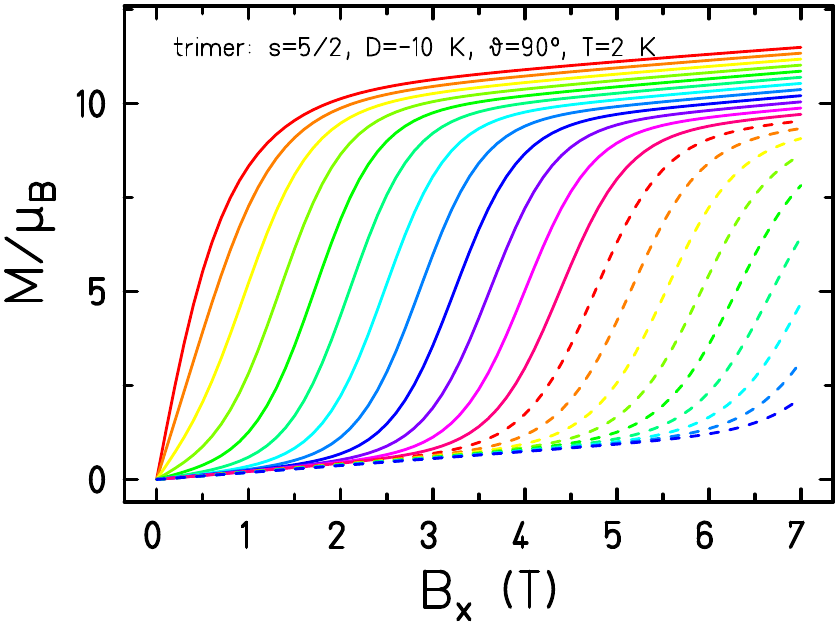}
\caption{Magnetization of the triangular spin arrangement (\figref{toroidal-f-c}): $s=5/2$, $D=-10$~K, curves for 
increasing antiferromagnetic coupling, $J=0, -0.1, -0.2, \dots -2.0$~K
from left to right. $B_x$ points along one of the easy axes.} 
\label{toroidal-f-x}
\end{figure}
%===================    figure  =================================

Figure \xref{toroidal-f-x} nicely demonstrates that the magnetization
may assume an S-shape, here shown for the triangle introduced in \figref{toroidal-f-c}. 
The curves show the magnetization for three
spins $s=5/2$ for a strong easy-axis anisotropy $D=-10$~K as function of 
increasing antiferromagnetic coupling, $J=0, -0.1, -0.2, \dots -2.0$~K
from left to right. $B_x$ points along one of the easy axes.
This figure does not change if the easy axes are oriented in toroidal
fashion, \figref{toroidal-f-c}(a) or are collectively rotated and 
point along radial directions, \figref{toroidal-f-c}(b).

%===================    figure   =================================
\begin{figure}[ht!]
\centering
\includegraphics*[clip,height=30mm]{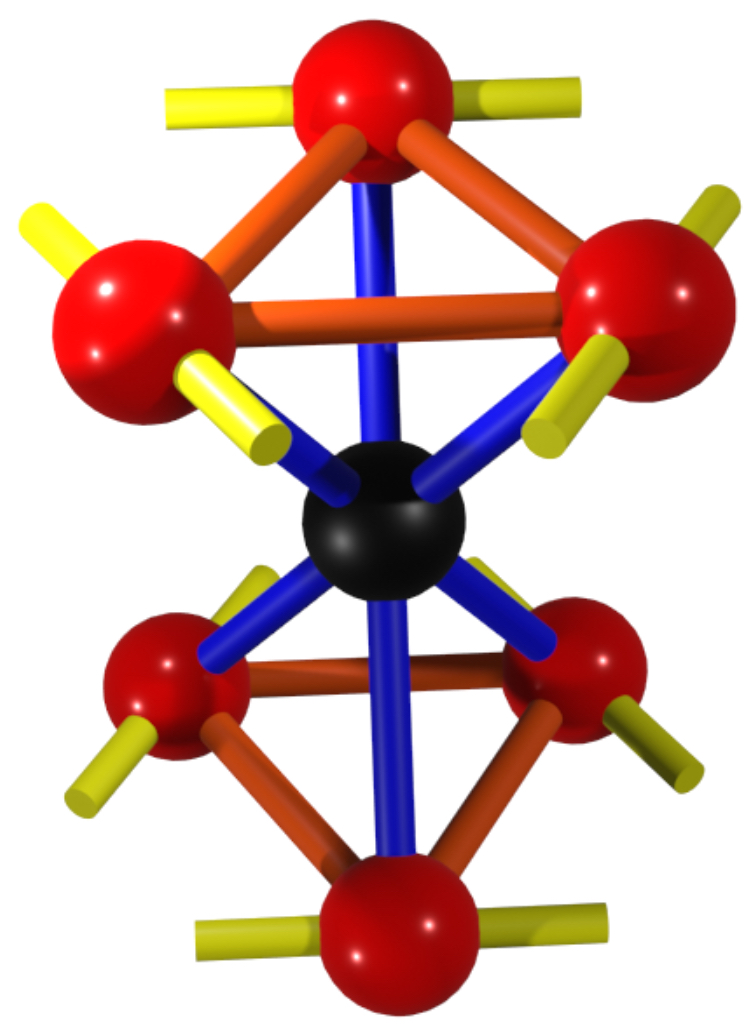}
\includegraphics*[clip,height=30mm]{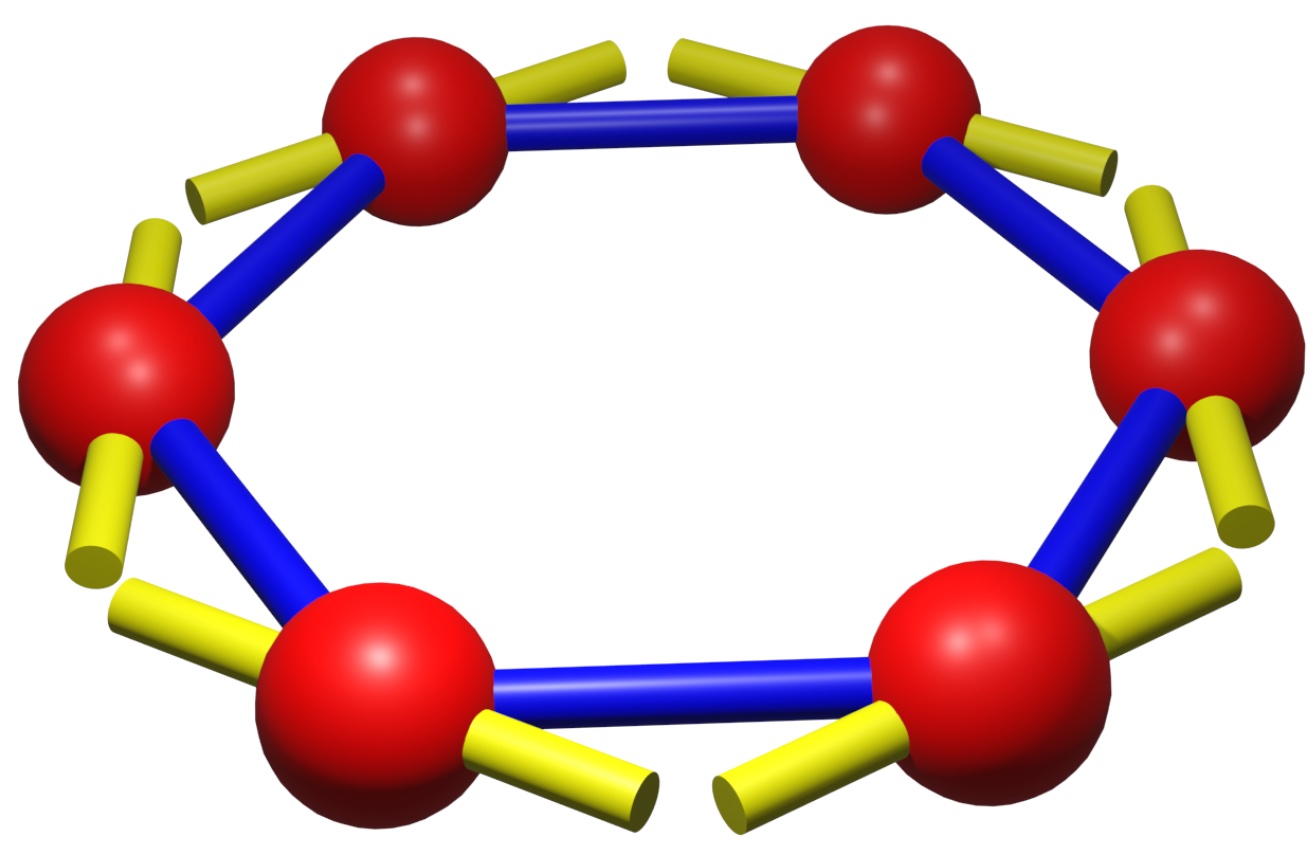}
\caption{Structures of an hourglass molecule (l.h.s.) and a
  hexagonal ring (r.h.s.). Yellow sticks represent the easy
  axes; the red and blue connections represent $J_1$ and $J_2$
  in the hourglass, whereas the blue connections represent $J$
  for the ring.} 
\label{toroidal-f-g}
\end{figure}
%===================    figure  =================================

In general, the situation is much more involved. It seems that
one needs a certain strength of exchange interaction compared to
the easy-axis anisotropy in order to obtain S-shape magnetization curves. 
We provide two examples along this
line: an hourglass-like spin systems that might stand for
Dy$_6$Cr and similar compounds \cite{VSL:NC17,ABV:EJIC21} 
and a hexagonal ring. For simplicity,
the easy axes are aligned in toroidal fashion in a plane,
see \figref{toroidal-f-g}. When looking at these structures 
one should keep in mind that the magnetization as well as other 
magnetic properties do not change, when all anisotropy tensors are 
rotated by a common angle of $90^\circ$ about the field axis. 
The toroidal moment could collapse to zero under such a transformation.

%===================    figure   =================================
\begin{figure}[ht!]
\centering
\includegraphics*[clip,width=0.85\columnwidth]{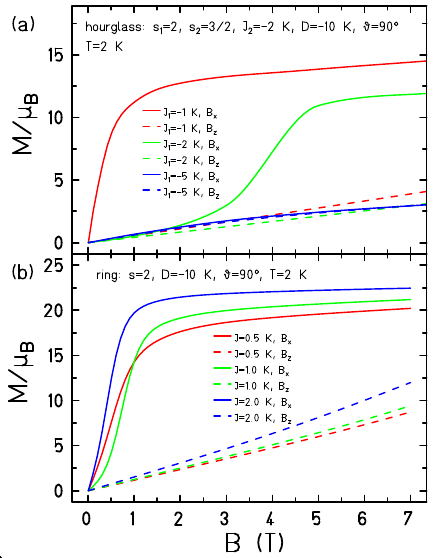}
\caption{Magnetization curves for various parameter sets as
  function of applied field along $x$- or $z$-direction for (top)
  an hourglass molecule and (bottom) a hexagonal ring. $z$ denotes
  the direction perpendicular to the triangular or hexagonal
  planes whereas $x$ is in plane.} 
\label{toroidal-f-h}
\end{figure}
%===================    figure  =================================

Figure~\xref{toroidal-f-h} shows several magnetization curves
along $x$- or $z$-direction for various parameter sets for an
hourglass molecule and a hexagonal ring. Depending on parameters
and field direction the curves might resemble an S-shape or
not, compare \cite{ABV:EJIC21} for similar experimental curves. 
The same also holds for the powder average which again
would not change if all anisotropy axes in \figref{toroidal-f-g}
would be rotated by $90^\circ$ about the $z$-axis to point 
radially outwards.

Sometimes the S-shape is taken as signature of a non-magnetic ground state,
but this statement is too weak because a non-magnetic ground state dominates
the low-field magnetization only if it is separated by a non-negligible 
energy gap from magnetic states. Thus an S-shaped magnetization curve 
signals that the low-energy spectrum is populated with non-magnetic states
whereas magnetic states appear only above some energy gap.

%%%%%%%%%%%%%%%%%%%%%%%%%%%%%%%%%%%%%%%%%%%%%%%%%%%%%%%%%%%%%%%%%%%%%%%%
\subsection{Tunneling gap}
\label{sec-2-3}

The key problem of toroidal arrangements of easy axes (including
the af dimer discussed above) is the tunneling gap at the
avoided level crossing of the two lowest energy eigenstates at
$B=0$. This practically unavoidable property of many spin Hamiltonians, in
particular in case of non-collinear easy axes, not only leads to
a quantum tunneling of the magnetization
\cite{HHK:IC12,GHG:CCR15}, but also of the toroidal moment. In
view of the symmetry discussed above, the tunneling rates are
just the same and thus a major obstacle against bistability and
thus technological use. 

%===================    table   =================================
\begin{table}
\centering
\begin{tabular}{c|c|c|c|c}
$D$ & -1.0 & -2.0 & -4.0 & -8.0\\ 
\hline\hline
$s=1$ & 0.561553 & 0.372281 & 0.216990 & 0.116844\\
\hline
$s=3/2$ &  0.227998 & 0.087343 & 0.027536 & 0.007767\\
\hline
$s=2$ & 0.072088 & 0.015878 & 0.002738 & 0.000407\\
\hline
$s=5/2$ & 0.019653 & 0.002519 & 0.000239 & 0.000019\\
\hline
\end{tabular}
\caption{Tunneling gaps $\Delta$ for antiferromagnetic dimers with $J=-0.5$~K 
and $D$ as well as $s$ as given in the table. 
All quantities are provided in kelvin. For real materials the
accuracy of the gap is of course not better than for $J$ and $D$.}
\label{tab-2-3-a}
\end{table}
%===================    table   =================================

%===================    table   =================================
\begin{table}
\centering
\begin{tabular}{c|c|c|c|c}
$D$ & -1.0 & -2.0 & -4.0 & -8.0\\ 
\hline\hline
$s=1$ & 0.415911 & 0.161685 & 0.047158 & 0.013061\\
\hline
$s=3/2$ & 0 & 0 & 0 & 0\\
\hline
$s=2$ & 0.011954 & 0.001137 & 0.000074 & 0.000004\\
\hline
$s=5/2$ & 0 & 0 & 0 & 0\\
\hline
\end{tabular}
\caption{Tunneling gaps $\Delta$ for antiferromagnetic trimers with $J=-0.5$~K 
and $D$ as well as $s$ as given in the table. 
All quantities are provided in kelvin. For real materials the
accuracy of the gap is of course not better than for $J$ and $D$.}
\label{tab-2-3-b}
\end{table}
%===================    table   =================================

We provide tunneling gaps $\Delta$ for antiferromagnetic dimers as well as 
trimers with $J=-0.5$~K and $D$ as well as $s$ in tables \xref{tab-2-3-a}
and \xref{tab-2-3-b}, respectively.
A slight prospect is provided by the observation that the
tunneling gap shrinks with increasing easy axes anisotropy $D_i$
of the participating spins as well as with increasing spin quantum
number \cite{Pis:M22}, the latter being a good argument to use
dysprosium in such compounds. 
We conjecture that the major reason for this behavior is that
with increasing spin quantum number as well as with increasing D
the contribution of the single-ion anisotropy to the total
energy increases. 
Since single-ion anisotropy is a one-body operator this
increases the anisotropy relative to the Heisenberg
interaction and therefore decreases the entanglement between the
spins, which is due to the Heisenberg interaction, in the
zero-field split ground states. This way the system 
approaches the limit of independent, i.e., non-interacting spins,
for which there is no avoided level crossing.

But even with a very small tunneling gap or for Kramers systems
(in total odd number of spin 1/2), where $\Delta=0$, the ground state 
might be very susceptible to small transverse fields since the anisotropy axes 
are not collinear and ground states are thus superpositions 
of basis states with
various magnetic quantum numbers, compare investigations in, e.g., 
Refs.~\cite{GHG:CCR15,LMB:IC17,BFW:CEJ19,OSG:EJIC20,HyS:MC22}.

%%%%%%%%%%%%%%%%%%%%%%%%%%%%%%%%%%%%%%%%%%%%%%%%%%%%%%%%%%%%%%%%%%%%%%%%
\section{Interactions that foster toroidal moments}
\label{sec-3}

Under which circumstances is the concept of toroidal moments
useful? We are convinced that one needs terms 
in the Hamiltonian that break the
symmetry of the discussed collective rotations. There are (at least) two
options: The exchange interactions are also anisotropic,
due to contributions of, e.g., antisymmetric 
Dzyaloshinskii-Moriya interaction, dipolar interaction, as
well as anisotropic symmetric exchange, or the magnetic field
depends on the space coordinates and has got cyclic
character, for instance.

The following Hamiltonian contains both options
%--------------------------------------------------------
\begin{eqnarray}
\label{E-3-1}
\op{H}
&=&
-
2\;
\sum_{i<j}\;
\op{\vec{s}}_i \cdot \mathbf{J}_{ij}
\cdot \op{\vec{s}}_j
+
\sum_{i}\;
\op{\vec{s}}_i \cdot 
{\mathbf D}_i
\cdot \op{\vec{s}}_i
\\
&&+
\mu_B\,
\sum_{i}\;
g_i
 \vec{B}(\vec{r}_i)\cdot\op{\vec{s}}_i
\nonumber
\ .
\end{eqnarray}
%--------------------------------------------------------
Here $\mathbf{J}_{ij}$ is the 3 by 3 matrix of the anisotropic
exchange between spins at sites $i$ and $j$.

Anisotropic exchange comprises all kinds of
anisotropic interactions, among them symmetric anisotropic
exchange, e.g.\ with 4d or 5d elements \cite{HSB:IC14} or Kitaev
interactions \cite{Kit:AP06,JaK:PRL09,WLJ:PRB16,ZZG:PRB21}
as well as antisymmetric anisotropic exchange of 
Dzyaloshinskii-Moriya type
\cite{Dzy:JPCS58,Mor:PRL60,Mor:PR60} and generalizations thereof,
e.g.\ topological-chiral magnetic interactions
\cite{GHH:NC20,HoB:PRB20,SBL:PRB21}. The strength of such
interactions does of course depend on the symmetry of the
chemical structure of the considered molecules (including
ligands etc.).
However, many 4d or 5d ions that show anisotropic exchange are
effective doublets (effective spins $s=1/2$), and therefore do not possess 
single-ion anisotropy. 
The simplest anisotropic interaction is the
dipolar interaction which acts between all kinds of magnetic moments
and in addition to all other terms in the Hamiltonian.

One should, however, keep in mind that the thermal stability of
toroidal effects is limited by the strength of the respective
anisotropic interaction at work. For instance, if the
contribution of the dipolar interaction to the Hamiltonian
amounts to 1~K, then one cannot expect it to stabilize toroidal
states for temperatures above this scale, and such a system will
be similar to that without dipolar interactions for higher
temperatures. We demonstrate this behavior with the following
example of a toroidal hexagonal ring, see r.h.s.\ of
\figref{toroidal-f-g}. 

%===================    figure   =================================
\begin{figure}[H]
\centering
\includegraphics*[clip,width=0.85\columnwidth]{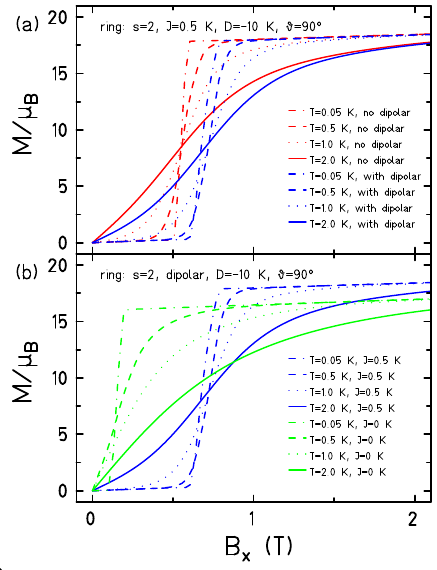}
\caption{Magnetization along the field direction of a spin ring with single spins $s=2$ and a toroidal arrangement
of easy axes as shown in \figref{toroidal-f-g}. (a) Comparison of magnetization curves 
for a ring with ferromagnetic nearest neighbor coupling of $J=0.5$ without (red) and 
with dipolar interactions (blue). (b) Comparison of magnetization curves 
for a ring with dipolar interactions with (blue) and without ferromagnetic 
nearest neighbor coupling of $J=0.5$ (green). The field is applied in the $xy$-plane defined by the toroidal anisotropy axes along one of these axes.
} 
\label{toroidal-f-i}
\end{figure}
%===================    figure  =================================

Figure~\xref{toroidal-f-i} compares three scenarios. Panel (a) compares
the magnetization curves for a ring with a ferromagnetic nearest neighbor 
coupling of $J=0.5$ without (red) and with dipolar interactions (blue). 
The dipolar interaction was taken to be realistic for a six-membered ring
such as in \cite{ULH:JACS12,LVG:DT19} 
($R=3.74$~\AA); it acts pairwise between all spins of the ring. 
One notices that the dipolar interaction indeed stabilizes the toroidal 
arrangement of the ground state since the field at which the magnetization 
jumps at low temperatures
is shifted to higher values. One also notices that this effect is 
weakened by higher temperatures; in particular at $T=2$~K it is almost gone.
It should be added here, that the dipolar interaction not necessarily 
stabilizes a toroidal moment; it may also counteract.

Panel (b) investigates how the dipolar interaction alone would perform
compared to a combined action of ferromagnetic and dipolar interaction.
The result is depicted by the green curves in \figref{toroidal-f-i}(b). 
They show that at least for the discussed example the magnetization is 
not stabilized against the magnetic field which leads us to conclude that
a combined action of isotropic and anisotropic exchange is preferential.

%===================    figure   =================================
\begin{figure}[ht!]
\centering
\includegraphics*[clip,width=0.75\columnwidth]{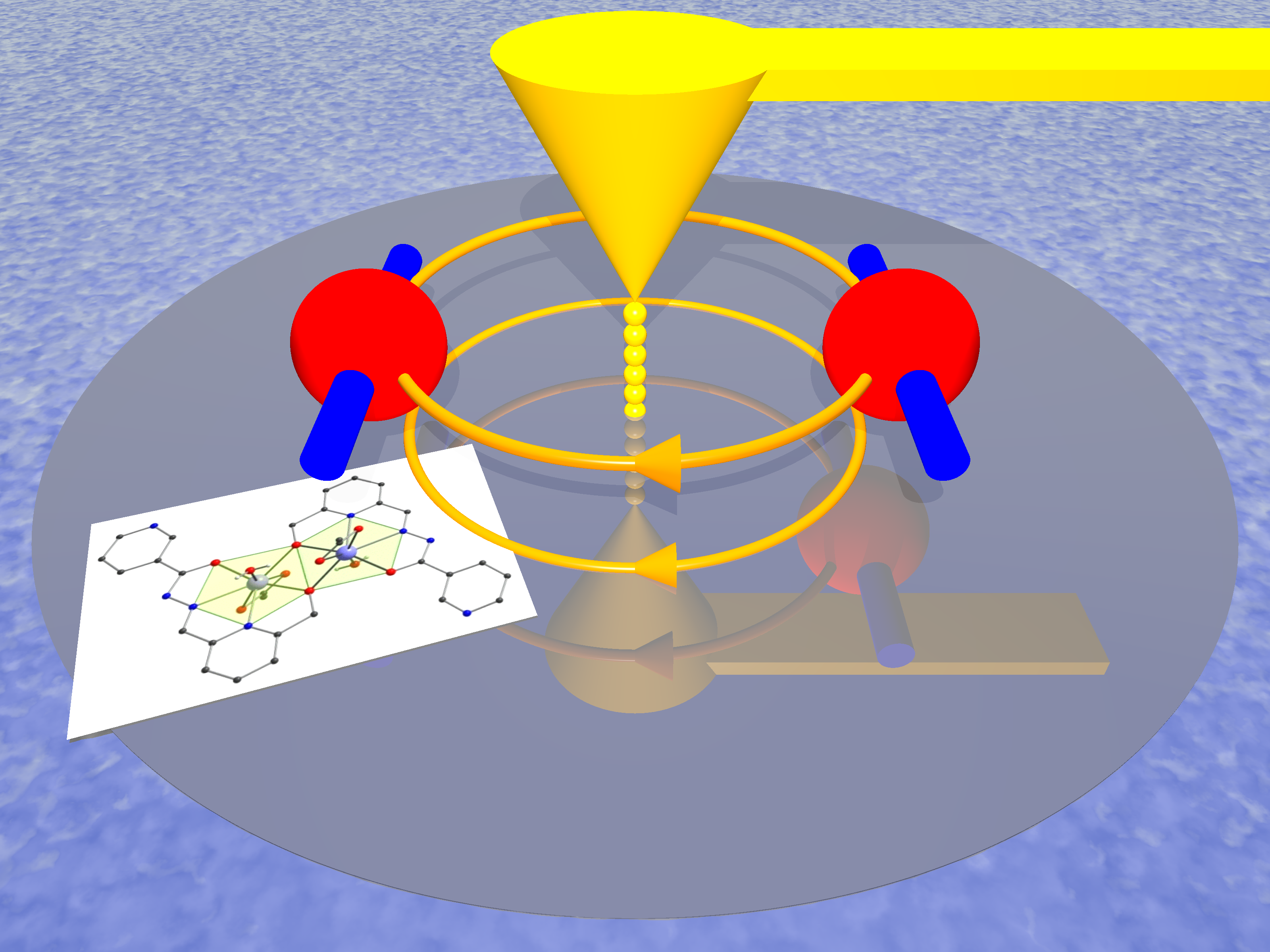}
\caption{Artistic view of a toroidal dimer whose toroidal quantum
states can be driven by the field of the tunneling current of
the scanning probe microscope. Unfortunately, the estimated
field of about $10^{-6}$~T for realistic tunnel currents
at the sites of the spins is too weak for practical use.} 
\label{toroidal-f-f}
\end{figure}
%===================    figure  =================================

Finally we would like to speculate about toroidal magnetic fields that 
would match toroidal states perfectly in the same way a homogeneous field
matches a collinear arrangement of ferromagnetically aligned spins. 
A perfectly suited magnetic field to pick up or initialize a toroidal moment
would be the field of a straight wire as for instance realized
by the tunneling current in a scanning tunneling microscope, see
artistic view in \figref{toroidal-f-f}. Unfortunately, such a tunneling current
generates a much too weak field of only about $10^{-6}$ Tesla 
for nowadays STMs \cite{LeA:PRB88}. 
One could however employ a magnetic tip that would initialize or pick up 
a toroidal moment by being placed above one of the magnetic ions as was
demonstrated in Ref.~\cite{BYW:SA18}.

\vspace*{3mm}

%%%%%%%%%%%%%%%%%%%%%%%%%%%%%%%%%%%%%%%%%%%%%%%%%%%%%%%%%%%%%%%%%%%%%%%%
\section{Discussion and conclusions}
\label{sec-dc}

There are three lessons to be learned from our investigation.

1. Even if a magnetic molecule possesses easy anisotropy axes 
that are arranged in a toroidal fashion its properties will not be related
to possible toroidal moments if the Hamiltonian consists dominantly of
Heisenberg exchange and single-ion anisotropy. The toroidal moment is a 
coincidence and does not influence the spectrum and thermal properties.

2. Toroidal moments might play a role if additional anisotropic exchange
enters the Hamiltonian \cite{HyS:MC22}. Then a symmetry transform as discussed 
above is not possible, and the toroidal moment might be stabilized by
the anisotropic exchange. In such a case one can hope to employ toroidal moments
for quantum devices. 

3. In physics one can usually estimate simple figures of merit by looking
at scales. Here, the thermal stability of the toroidal moment is given by the 
magnitude of the anisotropic interactions. If these terms sum up to less than a kelvin,
then the concept of a toroidal moment is useful below a kelvin and useless
above a kelvin. Thus, we need to search for materials where both the single-ion
anisotropy as well as the anisotropic exchange are large.

%%%%%%%%%%%%%%%%%%%%%%%%%%%%%%%%%%%%%%%%%%%%%%%%%%%%%%%%%%%%%%%%%%%%%%%%
\section*{Acknowledgment}

This work was supported by the Deutsche Forschungsgemeinschaft DFG
(355031190 (FOR~2692); 397300368 (SCHN~615/25-2)). 
Computing time at the Leibniz Center in Garching is gratefully
acknowledged.

%%%%%%%%%%%%%%%%%%%%%%%%%%%%%%%%%%%%%%%%%%%%%%%%%%%%%%%%%%%%%%%%%%%%%%%%
%\bibliographystyle{/home/schnack/tex/sty/revtex4-1/revtex4-1/bibtex/bst/revtex/apsrev4-2}
%\bibliography{/home/schnack/tex/bibtex/js-own.bib,/home/schnack/tex/bibtex/js-other.bib}

\begin{thebibliography}{67}%
\makeatletter
\providecommand \@ifxundefined [1]{%
 \@ifx{#1\undefined}
}%
\providecommand \@ifnum [1]{%
 \ifnum #1\expandafter \@firstoftwo
 \else \expandafter \@secondoftwo
 \fi
}%
\providecommand \@ifx [1]{%
 \ifx #1\expandafter \@firstoftwo
 \else \expandafter \@secondoftwo
 \fi
}%
\providecommand \natexlab [1]{#1}%
\providecommand \enquote  [1]{``#1''}%
\providecommand \bibnamefont  [1]{#1}%
\providecommand \bibfnamefont [1]{#1}%
\providecommand \citenamefont [1]{#1}%
\providecommand \href@noop [0]{\@secondoftwo}%
\providecommand \href [0]{\begingroup \@sanitize@url \@href}%
\providecommand \@href[1]{\@@startlink{#1}\@@href}%
\providecommand \@@href[1]{\endgroup#1\@@endlink}%
\providecommand \@sanitize@url [0]{\catcode `\\12\catcode `\$12\catcode
  `\&12\catcode `\#12\catcode `\^12\catcode `\_12\catcode `\%12\relax}%
\providecommand \@@startlink[1]{}%
\providecommand \@@endlink[0]{}%
\providecommand \url  [0]{\begingroup\@sanitize@url \@url }%
\providecommand \@url [1]{\endgroup\@href {#1}{\urlprefix }}%
\providecommand \urlprefix  [0]{URL }%
\providecommand \Eprint [0]{\href }%
\providecommand \doibase [0]{http://dx.doi.org/}%
\providecommand \selectlanguage [0]{\@gobble}%
\providecommand \bibinfo  [0]{\@secondoftwo}%
\providecommand \bibfield  [0]{\@secondoftwo}%
\providecommand \translation [1]{[#1]}%
\providecommand \BibitemOpen [0]{}%
\providecommand \bibitemStop [0]{}%
\providecommand \bibitemNoStop [0]{.\EOS\space}%
\providecommand \EOS [0]{\spacefactor3000\relax}%
\providecommand \BibitemShut  [1]{\csname bibitem#1\endcsname}%
\let\auto@bib@innerbib\@empty
%</preamble>
\bibitem [{\citenamefont {Gatteschi}\ \emph {et~al.}(2006)\citenamefont
  {Gatteschi}, \citenamefont {Sessoli},\ and\ \citenamefont
  {Villain}}]{GSV:2006}%
  \BibitemOpen
  \bibfield  {author} {\bibinfo {author} {\bibfnamefont {D.}~\bibnamefont
  {Gatteschi}}, \bibinfo {author} {\bibfnamefont {R.}~\bibnamefont {Sessoli}},
  \ and\ \bibinfo {author} {\bibfnamefont {J.}~\bibnamefont {Villain}},\
  }\href@noop {} {\emph {\bibinfo {title} {Molecular Nanomagnets}}},\
  Mesoscopic Physics and Nanotechnology\ (\bibinfo {address} {Oxford},\
  \bibinfo {year} {2006})\BibitemShut {NoStop}%
\bibitem [{\citenamefont {Blundell}(2007)}]{Blundell:CP07}%
  \BibitemOpen
  \bibfield  {author} {\bibinfo {author} {\bibfnamefont {S.~J.}\ \bibnamefont
  {Blundell}},\ }\enquote {\bibinfo {title} {Molecular magnets},}\ \href
  {\doibase 10.1080/00107510801967415} {\bibfield  {journal} {\bibinfo
  {journal} {Contemp. Phys.}\ }\textbf {\bibinfo {volume} {48}},\ \bibinfo
  {pages} {275} (\bibinfo {year} {2007})}\BibitemShut {NoStop}%
\bibitem [{\citenamefont {Furrer}\ and\ \citenamefont
  {Waldmann}(2013)}]{FuW:RMP13}%
  \BibitemOpen
  \bibfield  {author} {\bibinfo {author} {\bibfnamefont {A.}~\bibnamefont
  {Furrer}}\ and\ \bibinfo {author} {\bibfnamefont {O.}~\bibnamefont
  {Waldmann}},\ }\enquote {\bibinfo {title} {Magnetic cluster excitations},}\
  \href {http://link.aps.org/doi/10.1103/RevModPhys.85.367} {\bibfield
  {journal} {\bibinfo  {journal} {Rev. Mod. Phys.}\ }\textbf {\bibinfo {volume}
  {85}},\ \bibinfo {pages} {367} (\bibinfo {year} {2013})}\BibitemShut
  {NoStop}%
\bibitem [{\citenamefont {Liddle}\ and\ \citenamefont {van
  Slageren}(2015)}]{LvS:CSR15}%
  \BibitemOpen
  \bibfield  {author} {\bibinfo {author} {\bibfnamefont {S.~T.}\ \bibnamefont
  {Liddle}}\ and\ \bibinfo {author} {\bibfnamefont {J.}~\bibnamefont {van
  Slageren}},\ }\enquote {\bibinfo {title} {Improving f-element single molecule
  magnets},}\ \href {\doibase 10.1039/C5CS00222B} {\bibfield  {journal}
  {\bibinfo  {journal} {Chem. Soc. Rev.}\ }\textbf {\bibinfo {volume} {44}},\
  \bibinfo {pages} {6655} (\bibinfo {year} {2015})}\BibitemShut {NoStop}%
\bibitem [{\citenamefont {Schnack}(2019)}]{Sch:CP19}%
  \BibitemOpen
  \bibfield  {author} {\bibinfo {author} {\bibfnamefont {J.}~\bibnamefont
  {Schnack}},\ }\enquote {\bibinfo {title} {Large magnetic molecules and what
  we learn from them},}\ \href {\doibase 10.1080/00107514.2019.1615716}
  {\bibfield  {journal} {\bibinfo  {journal} {Contemp. Phys.}\ }\textbf
  {\bibinfo {volume} {60}},\ \bibinfo {pages} {127} (\bibinfo {year}
  {2019})}\BibitemShut {NoStop}%
\bibitem [{\citenamefont {Friedman}\ \emph {et~al.}(1996)\citenamefont
  {Friedman}, \citenamefont {Sarachik}, \citenamefont {Tejada},\ and\
  \citenamefont {Ziolo}}]{FST:PRL96}%
  \BibitemOpen
  \bibfield  {author} {\bibinfo {author} {\bibfnamefont {J.~R.}\ \bibnamefont
  {Friedman}}, \bibinfo {author} {\bibfnamefont {M.~P.}\ \bibnamefont
  {Sarachik}}, \bibinfo {author} {\bibfnamefont {J.}~\bibnamefont {Tejada}}, \
  and\ \bibinfo {author} {\bibfnamefont {R.}~\bibnamefont {Ziolo}},\ }\enquote
  {\bibinfo {title} {Macroscopic Measurement of Resonant Magnetization
  Tunneling in High-Spin Molecules},}\ \href {\doibase
  10.1103/PhysRevLett.76.3830} {\bibfield  {journal} {\bibinfo  {journal}
  {Phys. Rev. Lett.}\ }\textbf {\bibinfo {volume} {76}},\ \bibinfo {pages}
  {3830} (\bibinfo {year} {1996})}\BibitemShut {NoStop}%
\bibitem [{\citenamefont {Thomas}\ \emph {et~al.}(1996)\citenamefont {Thomas},
  \citenamefont {Lionti}, \citenamefont {Ballou}, \citenamefont {Gatteschi},
  \citenamefont {Sessoli},\ and\ \citenamefont {Barbara}}]{TLB:Nature96}%
  \BibitemOpen
  \bibfield  {author} {\bibinfo {author} {\bibfnamefont {L.}~\bibnamefont
  {Thomas}}, \bibinfo {author} {\bibfnamefont {F.}~\bibnamefont {Lionti}},
  \bibinfo {author} {\bibfnamefont {R.}~\bibnamefont {Ballou}}, \bibinfo
  {author} {\bibfnamefont {D.}~\bibnamefont {Gatteschi}}, \bibinfo {author}
  {\bibfnamefont {R.}~\bibnamefont {Sessoli}}, \ and\ \bibinfo {author}
  {\bibfnamefont {B.}~\bibnamefont {Barbara}},\ }\enquote {\bibinfo {title}
  {Macroscopic quantum tunnelling of magnetization in a single crystal of
  nanomagnets},}\ \href {http://dx.doi.org/10.1038/383145a0} {\bibfield
  {journal} {\bibinfo  {journal} {Nature}\ }\textbf {\bibinfo {volume} {383}},\
  \bibinfo {pages} {145} (\bibinfo {year} {1996})}\BibitemShut {NoStop}%
\bibitem [{\citenamefont {Goodwin}\ \emph {et~al.}(2017)\citenamefont
  {Goodwin}, \citenamefont {Ortu}, \citenamefont {Reta}, \citenamefont
  {Chilton},\ and\ \citenamefont {Mills}}]{GOR:N17}%
  \BibitemOpen
  \bibfield  {author} {\bibinfo {author} {\bibfnamefont {C.~A.~P.}\
  \bibnamefont {Goodwin}}, \bibinfo {author} {\bibfnamefont {F.}~\bibnamefont
  {Ortu}}, \bibinfo {author} {\bibfnamefont {D.}~\bibnamefont {Reta}}, \bibinfo
  {author} {\bibfnamefont {N.~F.}\ \bibnamefont {Chilton}}, \ and\ \bibinfo
  {author} {\bibfnamefont {D.~P.}\ \bibnamefont {Mills}},\ }\enquote {\bibinfo
  {title} {Molecular magnetic hysteresis at 60 kelvin in dysprosocenium},}\
  \href {\doibase 10.1038/nature23447} {\bibfield  {journal} {\bibinfo
  {journal} {Nature}\ }\textbf {\bibinfo {volume} {548}},\ \bibinfo {pages}
  {439} (\bibinfo {year} {2017})}\BibitemShut {NoStop}%
\bibitem [{\citenamefont {Randall~McClain}\ \emph {et~al.}(2018)\citenamefont
  {Randall~McClain}, \citenamefont {Gould}, \citenamefont {Chakarawet},
  \citenamefont {Teat}, \citenamefont {Groshens}, \citenamefont {Long},\ and\
  \citenamefont {Harvey}}]{RGC:CS18}%
  \BibitemOpen
  \bibfield  {author} {\bibinfo {author} {\bibfnamefont {K.}~\bibnamefont
  {Randall~McClain}}, \bibinfo {author} {\bibfnamefont {C.~A.}\ \bibnamefont
  {Gould}}, \bibinfo {author} {\bibfnamefont {K.}~\bibnamefont {Chakarawet}},
  \bibinfo {author} {\bibfnamefont {S.~J.}\ \bibnamefont {Teat}}, \bibinfo
  {author} {\bibfnamefont {T.~J.}\ \bibnamefont {Groshens}}, \bibinfo {author}
  {\bibfnamefont {J.~R.}\ \bibnamefont {Long}}, \ and\ \bibinfo {author}
  {\bibfnamefont {B.~G.}\ \bibnamefont {Harvey}},\ }\enquote {\bibinfo {title}
  {High-temperature magnetic blocking and magneto-structural correlations in a
  series of dysprosium(iii) metallocenium single-molecule magnets},}\ \href
  {\doibase 10.1039/C8SC03907K} {\bibfield  {journal} {\bibinfo  {journal}
  {Chem. Sci.}\ }\textbf {\bibinfo {volume} {9}},\ \bibinfo {pages} {8492}
  (\bibinfo {year} {2018})}\BibitemShut {NoStop}%
\bibitem [{\citenamefont {Guo}\ \emph {et~al.}(2018)\citenamefont {Guo},
  \citenamefont {Day}, \citenamefont {Chen}, \citenamefont {Tong},
  \citenamefont {Mansikkam{\"a}ki},\ and\ \citenamefont {Layfield}}]{GDC:S18}%
  \BibitemOpen
  \bibfield  {author} {\bibinfo {author} {\bibfnamefont {F.-S.}\ \bibnamefont
  {Guo}}, \bibinfo {author} {\bibfnamefont {B.~M.}\ \bibnamefont {Day}},
  \bibinfo {author} {\bibfnamefont {Y.-C.}\ \bibnamefont {Chen}}, \bibinfo
  {author} {\bibfnamefont {M.-L.}\ \bibnamefont {Tong}}, \bibinfo {author}
  {\bibfnamefont {A.}~\bibnamefont {Mansikkam{\"a}ki}}, \ and\ \bibinfo
  {author} {\bibfnamefont {R.~A.}\ \bibnamefont {Layfield}},\ }\enquote
  {\bibinfo {title} {Magnetic hysteresis up to 80 kelvin in a dysprosium
  metallocene single-molecule magnet},}\ \href {\doibase
  10.1126/science.aav0652} {\bibfield  {journal} {\bibinfo  {journal}
  {Science}\ }\textbf {\bibinfo {volume} {362}},\ \bibinfo {pages} {1400}
  (\bibinfo {year} {2018})}\BibitemShut {NoStop}%
\bibitem [{\citenamefont {Gould}\ \emph {et~al.}(2022)\citenamefont {Gould},
  \citenamefont {McClain}, \citenamefont {Reta}, \citenamefont {Kragskow},
  \citenamefont {Marchiori}, \citenamefont {Lachman}, \citenamefont {Choi},
  \citenamefont {Analytis}, \citenamefont {Britt}, \citenamefont {Chilton},
  \citenamefont {Harvey},\ and\ \citenamefont {Long}}]{GMR:S22}%
  \BibitemOpen
  \bibfield  {author} {\bibinfo {author} {\bibfnamefont {C.~A.}\ \bibnamefont
  {Gould}}, \bibinfo {author} {\bibfnamefont {K.~R.}\ \bibnamefont {McClain}},
  \bibinfo {author} {\bibfnamefont {D.}~\bibnamefont {Reta}}, \bibinfo {author}
  {\bibfnamefont {J.~G.~C.}\ \bibnamefont {Kragskow}}, \bibinfo {author}
  {\bibfnamefont {D.~A.}\ \bibnamefont {Marchiori}}, \bibinfo {author}
  {\bibfnamefont {E.}~\bibnamefont {Lachman}}, \bibinfo {author} {\bibfnamefont
  {E.-S.}\ \bibnamefont {Choi}}, \bibinfo {author} {\bibfnamefont {J.~G.}\
  \bibnamefont {Analytis}}, \bibinfo {author} {\bibfnamefont {R.~D.}\
  \bibnamefont {Britt}}, \bibinfo {author} {\bibfnamefont {N.~F.}\ \bibnamefont
  {Chilton}}, \bibinfo {author} {\bibfnamefont {B.~G.}\ \bibnamefont {Harvey}},
  \ and\ \bibinfo {author} {\bibfnamefont {J.~R.}\ \bibnamefont {Long}},\
  }\enquote {\bibinfo {title} {Ultrahard magnetism from mixed-valence
  dilanthanide complexes with metal-metal bonding},}\ \href {\doibase
  10.1126/science.abl5470} {\bibfield  {journal} {\bibinfo  {journal}
  {Science}\ }\textbf {\bibinfo {volume} {375}},\ \bibinfo {pages} {198}
  (\bibinfo {year} {2022})}\BibitemShut {NoStop}%
\bibitem [{\citenamefont {Irl\"ander}\ and\ \citenamefont
  {Schnack}(2020)}]{IrS:PRB20}%
  \BibitemOpen
  \bibfield  {author} {\bibinfo {author} {\bibfnamefont {K.}~\bibnamefont
  {Irl\"ander}}\ and\ \bibinfo {author} {\bibfnamefont {J.}~\bibnamefont
  {Schnack}},\ }\enquote {\bibinfo {title} {Spin-phonon interaction induces
  tunnel splitting in single-molecule magnets},}\ \href {\doibase
  10.1103/PhysRevB.102.054407} {\bibfield  {journal} {\bibinfo  {journal}
  {Phys. Rev. B}\ }\textbf {\bibinfo {volume} {102}},\ \bibinfo {pages}
  {054407} (\bibinfo {year} {2020})}\BibitemShut {NoStop}%
\bibitem [{\citenamefont {Hymas}\ and\ \citenamefont
  {Soncini}(2022)}]{HyS:MC22}%
  \BibitemOpen
  \bibfield  {author} {\bibinfo {author} {\bibfnamefont {K.}~\bibnamefont
  {Hymas}}\ and\ \bibinfo {author} {\bibfnamefont {A.}~\bibnamefont
  {Soncini}},\ }\enquote {\bibinfo {title} {The Role of Magnetic Dipole-Dipole
  Coupling in Quantum Single-Molecule Toroics},}\ \href {\doibase
  10.3390/magnetochemistry8050058} {\bibfield  {journal} {\bibinfo  {journal}
  {Magnetochemistry}\ }\textbf {\bibinfo {volume} {8}},\ \bibinfo {pages} {58}
  (\bibinfo {year} {2022})}\BibitemShut {NoStop}%
\bibitem [{\citenamefont {Ardavan}\ \emph {et~al.}(2007)\citenamefont
  {Ardavan}, \citenamefont {Rival}, \citenamefont {Morton}, \citenamefont
  {Blundell}, \citenamefont {Tyryshkin}, \citenamefont {Timco},\ and\
  \citenamefont {Winpenny}}]{ARM:PRL07}%
  \BibitemOpen
  \bibfield  {author} {\bibinfo {author} {\bibfnamefont {A.}~\bibnamefont
  {Ardavan}}, \bibinfo {author} {\bibfnamefont {O.}~\bibnamefont {Rival}},
  \bibinfo {author} {\bibfnamefont {J.~J.~L.}\ \bibnamefont {Morton}}, \bibinfo
  {author} {\bibfnamefont {S.~J.}\ \bibnamefont {Blundell}}, \bibinfo {author}
  {\bibfnamefont {A.~M.}\ \bibnamefont {Tyryshkin}}, \bibinfo {author}
  {\bibfnamefont {G.~A.}\ \bibnamefont {Timco}}, \ and\ \bibinfo {author}
  {\bibfnamefont {R.~E.~P.}\ \bibnamefont {Winpenny}},\ }\enquote {\bibinfo
  {title} {Will Spin-Relaxation Times in Molecular Magnets Permit Quantum
  Information Processing?}}\ \href
  {http://link.aps.org/abstract/PRL/v98/e057201} {\bibfield  {journal}
  {\bibinfo  {journal} {Phys. Rev. Lett.}\ }\textbf {\bibinfo {volume} {98}},\
  \bibinfo {pages} {057201} (\bibinfo {year} {2007})}\BibitemShut {NoStop}%
\bibitem [{\citenamefont {Wernsdorfer}(2007)}]{Wer:NM07}%
  \BibitemOpen
  \bibfield  {author} {\bibinfo {author} {\bibfnamefont {W.}~\bibnamefont
  {Wernsdorfer}},\ }\enquote {\bibinfo {title} {A long-lasting phase},}\ \href
  {\doibase 10.1038/nmat1852} {\bibfield  {journal} {\bibinfo  {journal}
  {Nature Materials}\ }\textbf {\bibinfo {volume} {6}},\ \bibinfo {pages} {174}
  (\bibinfo {year} {2007})}\BibitemShut {NoStop}%
\bibitem [{\citenamefont {Kaminski}\ \emph {et~al.}(2014)\citenamefont
  {Kaminski}, \citenamefont {Webber}, \citenamefont {Wedge}, \citenamefont
  {Liu}, \citenamefont {Timco}, \citenamefont {Vitorica-Yrezabal},
  \citenamefont {McInnes}, \citenamefont {Winpenny},\ and\ \citenamefont
  {Ardavan}}]{KWW:PRB14}%
  \BibitemOpen
  \bibfield  {author} {\bibinfo {author} {\bibfnamefont {D.}~\bibnamefont
  {Kaminski}}, \bibinfo {author} {\bibfnamefont {A.~L.}\ \bibnamefont
  {Webber}}, \bibinfo {author} {\bibfnamefont {C.~J.}\ \bibnamefont {Wedge}},
  \bibinfo {author} {\bibfnamefont {J.}~\bibnamefont {Liu}}, \bibinfo {author}
  {\bibfnamefont {G.~A.}\ \bibnamefont {Timco}}, \bibinfo {author}
  {\bibfnamefont {I.~n.~J.}\ \bibnamefont {Vitorica-Yrezabal}}, \bibinfo
  {author} {\bibfnamefont {E.~J.~L.}\ \bibnamefont {McInnes}}, \bibinfo
  {author} {\bibfnamefont {R.~E.~P.}\ \bibnamefont {Winpenny}}, \ and\ \bibinfo
  {author} {\bibfnamefont {A.}~\bibnamefont {Ardavan}},\ }\enquote {\bibinfo
  {title} {Quantum spin coherence in halogen-modified {Cr$_{7}$Ni} molecular
  nanomagnets},}\ \href {\doibase 10.1103/PhysRevB.90.184419} {\bibfield
  {journal} {\bibinfo  {journal} {Phys. Rev. B}\ }\textbf {\bibinfo {volume}
  {90}},\ \bibinfo {pages} {184419} (\bibinfo {year} {2014})}\BibitemShut
  {NoStop}%
\bibitem [{\citenamefont {Shiddiq}\ \emph {et~al.}(2016)\citenamefont
  {Shiddiq}, \citenamefont {Komijani}, \citenamefont {Duan}, \citenamefont
  {Gaita-Ari{\~n}o}, \citenamefont {Coronado},\ and\ \citenamefont
  {Hill}}]{SKD:N16}%
  \BibitemOpen
  \bibfield  {author} {\bibinfo {author} {\bibfnamefont {M.}~\bibnamefont
  {Shiddiq}}, \bibinfo {author} {\bibfnamefont {D.}~\bibnamefont {Komijani}},
  \bibinfo {author} {\bibfnamefont {Y.}~\bibnamefont {Duan}}, \bibinfo {author}
  {\bibfnamefont {A.}~\bibnamefont {Gaita-Ari{\~n}o}}, \bibinfo {author}
  {\bibfnamefont {E.}~\bibnamefont {Coronado}}, \ and\ \bibinfo {author}
  {\bibfnamefont {S.}~\bibnamefont {Hill}},\ }\enquote {\bibinfo {title}
  {Enhancing coherence in molecular spin qubits via atomic clock
  transitions},}\ \href {https://doi.org/10.1038/nature16984} {\bibfield
  {journal} {\bibinfo  {journal} {Nature}\ }\textbf {\bibinfo {volume} {531}},\
  \bibinfo {pages} {348} (\bibinfo {year} {2016})}\BibitemShut {NoStop}%
\bibitem [{\citenamefont {Godfrin}\ \emph {et~al.}(2017)\citenamefont
  {Godfrin}, \citenamefont {Ferhat}, \citenamefont {Ballou}, \citenamefont
  {Klyatskaya}, \citenamefont {Ruben}, \citenamefont {Wernsdorfer},\ and\
  \citenamefont {Balestro}}]{GFB:PRL17}%
  \BibitemOpen
  \bibfield  {author} {\bibinfo {author} {\bibfnamefont {C.}~\bibnamefont
  {Godfrin}}, \bibinfo {author} {\bibfnamefont {A.}~\bibnamefont {Ferhat}},
  \bibinfo {author} {\bibfnamefont {R.}~\bibnamefont {Ballou}}, \bibinfo
  {author} {\bibfnamefont {S.}~\bibnamefont {Klyatskaya}}, \bibinfo {author}
  {\bibfnamefont {M.}~\bibnamefont {Ruben}}, \bibinfo {author} {\bibfnamefont
  {W.}~\bibnamefont {Wernsdorfer}}, \ and\ \bibinfo {author} {\bibfnamefont
  {F.}~\bibnamefont {Balestro}},\ }\enquote {\bibinfo {title} {Operating
  Quantum States in Single Magnetic Molecules: Implementation of {G}rover's
  Quantum Algorithm},}\ \href {\doibase 10.1103/PhysRevLett.119.187702}
  {\bibfield  {journal} {\bibinfo  {journal} {Phys. Rev. Lett.}\ }\textbf
  {\bibinfo {volume} {119}},\ \bibinfo {pages} {187702} (\bibinfo {year}
  {2017})}\BibitemShut {NoStop}%
\bibitem [{\citenamefont {Gaita-Ari{\~n}o}\ \emph {et~al.}(2019)\citenamefont
  {Gaita-Ari{\~n}o}, \citenamefont {Luis}, \citenamefont {Hill},\ and\
  \citenamefont {Coronado}}]{GLH:NC19}%
  \BibitemOpen
  \bibfield  {author} {\bibinfo {author} {\bibfnamefont {A.}~\bibnamefont
  {Gaita-Ari{\~n}o}}, \bibinfo {author} {\bibfnamefont {F.}~\bibnamefont
  {Luis}}, \bibinfo {author} {\bibfnamefont {S.}~\bibnamefont {Hill}}, \ and\
  \bibinfo {author} {\bibfnamefont {E.}~\bibnamefont {Coronado}},\ }\enquote
  {\bibinfo {title} {Molecular spins for quantum computation},}\ \href
  {\doibase 10.1038/s41557-019-0232-y} {\bibfield  {journal} {\bibinfo
  {journal} {Nature Chemistry}\ }\textbf {\bibinfo {volume} {11}},\ \bibinfo
  {pages} {301} (\bibinfo {year} {2019})}\BibitemShut {NoStop}%
\bibitem [{\citenamefont {Atzori}\ and\ \citenamefont
  {Sessoli}(2019)}]{AtS:JACS19}%
  \BibitemOpen
  \bibfield  {author} {\bibinfo {author} {\bibfnamefont {M.}~\bibnamefont
  {Atzori}}\ and\ \bibinfo {author} {\bibfnamefont {R.}~\bibnamefont
  {Sessoli}},\ }\enquote {\bibinfo {title} {The Second Quantum Revolution: Role
  and Challenges of Molecular Chemistry},}\ \href {\doibase
  10.1021/jacs.9b00984} {\bibfield  {journal} {\bibinfo  {journal} {J. Am.
  Chem. Soc.}\ }\textbf {\bibinfo {volume} {141}},\ \bibinfo {pages} {11339}
  (\bibinfo {year} {2019})}\BibitemShut {NoStop}%
\bibitem [{\citenamefont {Collett}\ \emph {et~al.}(2020)\citenamefont
  {Collett}, \citenamefont {Santini}, \citenamefont {Carretta},\ and\
  \citenamefont {Friedman}}]{CSC:PRR20}%
  \BibitemOpen
  \bibfield  {author} {\bibinfo {author} {\bibfnamefont {C.~A.}\ \bibnamefont
  {Collett}}, \bibinfo {author} {\bibfnamefont {P.}~\bibnamefont {Santini}},
  \bibinfo {author} {\bibfnamefont {S.}~\bibnamefont {Carretta}}, \ and\
  \bibinfo {author} {\bibfnamefont {J.~R.}\ \bibnamefont {Friedman}},\
  }\enquote {\bibinfo {title} {Constructing clock-transition-based two-qubit
  gates from dimers of molecular nanomagnets},}\ \href {\doibase
  10.1103/PhysRevResearch.2.032037} {\bibfield  {journal} {\bibinfo  {journal}
  {Phys. Rev. Research}\ }\textbf {\bibinfo {volume} {2}},\ \bibinfo {pages}
  {032037} (\bibinfo {year} {2020})}\BibitemShut {NoStop}%
\bibitem [{\citenamefont {Carretta}\ \emph {et~al.}(2021)\citenamefont
  {Carretta}, \citenamefont {Zueco}, \citenamefont {Chiesa}, \citenamefont
  {Gomez-Leon},\ and\ \citenamefont {Luis}}]{CZC:APL21}%
  \BibitemOpen
  \bibfield  {author} {\bibinfo {author} {\bibfnamefont {S.}~\bibnamefont
  {Carretta}}, \bibinfo {author} {\bibfnamefont {D.}~\bibnamefont {Zueco}},
  \bibinfo {author} {\bibfnamefont {A.}~\bibnamefont {Chiesa}}, \bibinfo
  {author} {\bibfnamefont {A.}~\bibnamefont {Gomez-Leon}}, \ and\ \bibinfo
  {author} {\bibfnamefont {F.}~\bibnamefont {Luis}},\ }\enquote {\bibinfo
  {title} {A perspective on scaling up quantum computation with molecular
  spins},}\ \href {\doibase 10.1063/5.0053378} {\bibfield  {journal} {\bibinfo
  {journal} {Appl. Phys. Lett.}\ }\textbf {\bibinfo {volume} {118}},\ \bibinfo
  {pages} {240501} (\bibinfo {year} {2021})}\BibitemShut {NoStop}%
\bibitem [{\citenamefont {Petiziol}\ \emph {et~al.}(2021)\citenamefont
  {Petiziol}, \citenamefont {Chiesa}, \citenamefont {Wimberger}, \citenamefont
  {Santini},\ and\ \citenamefont {Carretta}}]{PCW:npjQI21}%
  \BibitemOpen
  \bibfield  {author} {\bibinfo {author} {\bibfnamefont {F.}~\bibnamefont
  {Petiziol}}, \bibinfo {author} {\bibfnamefont {A.}~\bibnamefont {Chiesa}},
  \bibinfo {author} {\bibfnamefont {S.}~\bibnamefont {Wimberger}}, \bibinfo
  {author} {\bibfnamefont {P.}~\bibnamefont {Santini}}, \ and\ \bibinfo
  {author} {\bibfnamefont {S.}~\bibnamefont {Carretta}},\ }\enquote {\bibinfo
  {title} {Counteracting dephasing in Molecular Nanomagnets by optimized qudit
  encodings},}\ \href {\doibase 10.1038/s41534-021-00466-3} {\bibfield
  {journal} {\bibinfo  {journal} {npj Quantum Information}\ }\textbf {\bibinfo
  {volume} {7}},\ \bibinfo {pages} {133} (\bibinfo {year} {2021})}\BibitemShut
  {NoStop}%
\bibitem [{\citenamefont {Liu}\ \emph {et~al.}(2021)\citenamefont {Liu},
  \citenamefont {Mrozek}, \citenamefont {Ullah}, \citenamefont {Duan},
  \citenamefont {Baldov{\'\i}}, \citenamefont {Coronado}, \citenamefont
  {Gaita-Ari{\~n}o},\ and\ \citenamefont {Ardavan}}]{LMU:NP21}%
  \BibitemOpen
  \bibfield  {author} {\bibinfo {author} {\bibfnamefont {J.}~\bibnamefont
  {Liu}}, \bibinfo {author} {\bibfnamefont {J.}~\bibnamefont {Mrozek}},
  \bibinfo {author} {\bibfnamefont {A.}~\bibnamefont {Ullah}}, \bibinfo
  {author} {\bibfnamefont {Y.}~\bibnamefont {Duan}}, \bibinfo {author}
  {\bibfnamefont {J.}~\bibnamefont {Baldov{\'\i}}}, \bibinfo {author}
  {\bibfnamefont {E.}~\bibnamefont {Coronado}}, \bibinfo {author}
  {\bibfnamefont {A.}~\bibnamefont {Gaita-Ari{\~n}o}}, \ and\ \bibinfo {author}
  {\bibfnamefont {A.}~\bibnamefont {Ardavan}},\ }\enquote {\bibinfo {title}
  {Quantum coherent spin--electric control in a molecular nanomagnet at clock
  transitions},}\ \href {\doibase 10.1038/s41567-021-01355-4} {\bibfield
  {journal} {\bibinfo  {journal} {Nature Physics}\ }\textbf {\bibinfo {volume}
  {17}},\ \bibinfo {pages} {1205} (\bibinfo {year} {2021})}\BibitemShut
  {NoStop}%
\bibitem [{\citenamefont {Chiesa}\ \emph {et~al.}(2022)\citenamefont {Chiesa},
  \citenamefont {Petiziol}, \citenamefont {Chizzini}, \citenamefont {Santini},\
  and\ \citenamefont {Carretta}}]{CPC:JPCL22}%
  \BibitemOpen
  \bibfield  {author} {\bibinfo {author} {\bibfnamefont {A.}~\bibnamefont
  {Chiesa}}, \bibinfo {author} {\bibfnamefont {F.}~\bibnamefont {Petiziol}},
  \bibinfo {author} {\bibfnamefont {M.}~\bibnamefont {Chizzini}}, \bibinfo
  {author} {\bibfnamefont {P.}~\bibnamefont {Santini}}, \ and\ \bibinfo
  {author} {\bibfnamefont {S.}~\bibnamefont {Carretta}},\ }\enquote {\bibinfo
  {title} {Theoretical Design of Optimal Molecular Qudits for Quantum Error
  Correction},}\ \href {\doibase 10.1021/acs.jpclett.2c01602} {\bibfield
  {journal} {\bibinfo  {journal} {J. Phys. Chem. Lett.}\ }\textbf {\bibinfo
  {volume} {13}},\ \bibinfo {pages} {6468} (\bibinfo {year}
  {2022})}\BibitemShut {NoStop}%
\bibitem [{\citenamefont {Tang}\ \emph {et~al.}(2006)\citenamefont {Tang},
  \citenamefont {Hewitt}, \citenamefont {Madhu}, \citenamefont {Chastanet},
  \citenamefont {Wernsdorfer}, \citenamefont {Anson}, \citenamefont {Benelli},
  \citenamefont {Sessoli},\ and\ \citenamefont {Powell}}]{THM:ACIE06}%
  \BibitemOpen
  \bibfield  {author} {\bibinfo {author} {\bibfnamefont {J.}~\bibnamefont
  {Tang}}, \bibinfo {author} {\bibfnamefont {I.}~\bibnamefont {Hewitt}},
  \bibinfo {author} {\bibfnamefont {N.~T.}\ \bibnamefont {Madhu}}, \bibinfo
  {author} {\bibfnamefont {G.}~\bibnamefont {Chastanet}}, \bibinfo {author}
  {\bibfnamefont {W.}~\bibnamefont {Wernsdorfer}}, \bibinfo {author}
  {\bibfnamefont {C.~E.}\ \bibnamefont {Anson}}, \bibinfo {author}
  {\bibfnamefont {C.}~\bibnamefont {Benelli}}, \bibinfo {author} {\bibfnamefont
  {R.}~\bibnamefont {Sessoli}}, \ and\ \bibinfo {author} {\bibfnamefont
  {A.~K.}\ \bibnamefont {Powell}},\ }\enquote {\bibinfo {title} {Dysprosium
  Triangles Showing Single-Molecule Magnet Behavior of Thermally Excited Spin
  States},}\ \href {\doibase https://doi.org/10.1002/anie.200503564} {\bibfield
   {journal} {\bibinfo  {journal} {Angew. Chem. Int. Ed.}\ }\textbf {\bibinfo
  {volume} {45}},\ \bibinfo {pages} {1729} (\bibinfo {year}
  {2006})}\BibitemShut {NoStop}%
\bibitem [{\citenamefont {van Aken}\ \emph {et~al.}(2007)\citenamefont {van
  Aken}, \citenamefont {Rivera}, \citenamefont {Schmid},\ and\ \citenamefont
  {Fiebig}}]{ARS:N07}%
  \BibitemOpen
  \bibfield  {author} {\bibinfo {author} {\bibfnamefont {B.~B.}\ \bibnamefont
  {van Aken}}, \bibinfo {author} {\bibfnamefont {J.-P.}\ \bibnamefont
  {Rivera}}, \bibinfo {author} {\bibfnamefont {H.}~\bibnamefont {Schmid}}, \
  and\ \bibinfo {author} {\bibfnamefont {M.}~\bibnamefont {Fiebig}},\ }\enquote
  {\bibinfo {title} {Observation of ferrotoroidic domains},}\ \href
  {http://dx.doi.org/10.1038/nature06139} {\bibfield  {journal} {\bibinfo
  {journal} {Nature}\ }\textbf {\bibinfo {volume} {449}},\ \bibinfo {pages}
  {702} (\bibinfo {year} {2007})}\BibitemShut {NoStop}%
\bibitem [{\citenamefont {Soncini}\ and\ \citenamefont
  {Chibotaru}(2008)}]{SoC:PRB08}%
  \BibitemOpen
  \bibfield  {author} {\bibinfo {author} {\bibfnamefont {A.}~\bibnamefont
  {Soncini}}\ and\ \bibinfo {author} {\bibfnamefont {L.~F.}\ \bibnamefont
  {Chibotaru}},\ }\enquote {\bibinfo {title} {Toroidal magnetic states in
  molecular wheels: Interplay between isotropic exchange interactions and local
  magnetic anisotropy},}\ \href {\doibase 10.1103/PhysRevB.77.220406}
  {\bibfield  {journal} {\bibinfo  {journal} {Phys. Rev. B}\ }\textbf {\bibinfo
  {volume} {77}},\ \bibinfo {pages} {220406} (\bibinfo {year}
  {2008})}\BibitemShut {NoStop}%
\bibitem [{\citenamefont {Spaldin}\ \emph {et~al.}(2008)\citenamefont
  {Spaldin}, \citenamefont {Fiebig},\ and\ \citenamefont
  {Mostovoy}}]{SFM:JPCM08}%
  \BibitemOpen
  \bibfield  {author} {\bibinfo {author} {\bibfnamefont {N.~A.}\ \bibnamefont
  {Spaldin}}, \bibinfo {author} {\bibfnamefont {M.}~\bibnamefont {Fiebig}}, \
  and\ \bibinfo {author} {\bibfnamefont {M.}~\bibnamefont {Mostovoy}},\
  }\enquote {\bibinfo {title} {The toroidal moment in condensed-matter physics
  and its relation to the magnetoelectric effect},}\ \href
  {http://stacks.iop.org/0953-8984/20/i=43/a=434203} {\bibfield  {journal}
  {\bibinfo  {journal} {J. Phys.: Condens. Matter}\ }\textbf {\bibinfo {volume}
  {20}},\ \bibinfo {pages} {434203} (\bibinfo {year} {2008})}\BibitemShut
  {NoStop}%
\bibitem [{\citenamefont {Ungur}\ \emph {et~al.}(2012)\citenamefont {Ungur},
  \citenamefont {Langley}, \citenamefont {Hooper}, \citenamefont {Moubaraki},
  \citenamefont {Brechin}, \citenamefont {Murray},\ and\ \citenamefont
  {Chibotaru}}]{ULH:JACS12}%
  \BibitemOpen
  \bibfield  {author} {\bibinfo {author} {\bibfnamefont {L.}~\bibnamefont
  {Ungur}}, \bibinfo {author} {\bibfnamefont {S.~K.}\ \bibnamefont {Langley}},
  \bibinfo {author} {\bibfnamefont {T.~N.}\ \bibnamefont {Hooper}}, \bibinfo
  {author} {\bibfnamefont {B.}~\bibnamefont {Moubaraki}}, \bibinfo {author}
  {\bibfnamefont {E.~K.}\ \bibnamefont {Brechin}}, \bibinfo {author}
  {\bibfnamefont {K.~S.}\ \bibnamefont {Murray}}, \ and\ \bibinfo {author}
  {\bibfnamefont {L.~F.}\ \bibnamefont {Chibotaru}},\ }\enquote {\bibinfo
  {title} {Net Toroidal Magnetic Moment in the Ground State of a
  {Dy6}-Triethanolamine Ring},}\ \href {\doibase 10.1021/ja309211d} {\bibfield
  {journal} {\bibinfo  {journal} {J. Am. Chem. Soc.}\ }\textbf {\bibinfo
  {volume} {134}},\ \bibinfo {pages} {18554} (\bibinfo {year}
  {2012})}\BibitemShut {NoStop}%
\bibitem [{\citenamefont {Wang}\ \emph {et~al.}(2012)\citenamefont {Wang},
  \citenamefont {Shi}, \citenamefont {Li}, \citenamefont {Song}, \citenamefont
  {Fang}, \citenamefont {Lan}, \citenamefont {Powell}, \citenamefont
  {Wernsdorfer}, \citenamefont {Ungur}, \citenamefont {Chibotaru},
  \citenamefont {Shen},\ and\ \citenamefont {Cheng}}]{WSL:CS12}%
  \BibitemOpen
  \bibfield  {author} {\bibinfo {author} {\bibfnamefont {Y.-X.}\ \bibnamefont
  {Wang}}, \bibinfo {author} {\bibfnamefont {W.}~\bibnamefont {Shi}}, \bibinfo
  {author} {\bibfnamefont {H.}~\bibnamefont {Li}}, \bibinfo {author}
  {\bibfnamefont {Y.}~\bibnamefont {Song}}, \bibinfo {author} {\bibfnamefont
  {L.}~\bibnamefont {Fang}}, \bibinfo {author} {\bibfnamefont {Y.}~\bibnamefont
  {Lan}}, \bibinfo {author} {\bibfnamefont {A.~K.}\ \bibnamefont {Powell}},
  \bibinfo {author} {\bibfnamefont {W.}~\bibnamefont {Wernsdorfer}}, \bibinfo
  {author} {\bibfnamefont {L.}~\bibnamefont {Ungur}}, \bibinfo {author}
  {\bibfnamefont {L.~F.}\ \bibnamefont {Chibotaru}}, \bibinfo {author}
  {\bibfnamefont {M.}~\bibnamefont {Shen}}, \ and\ \bibinfo {author}
  {\bibfnamefont {P.}~\bibnamefont {Cheng}},\ }\enquote {\bibinfo {title} {A
  single-molecule magnet assembly exhibiting a dielectric transition at 470
  K},}\ \href {\doibase 10.1039/C2SC21023A} {\bibfield  {journal} {\bibinfo
  {journal} {Chem. Sci.}\ }\textbf {\bibinfo {volume} {3}},\ \bibinfo {pages}
  {3366} (\bibinfo {year} {2012})}\BibitemShut {NoStop}%
\bibitem [{\citenamefont {Xue}\ \emph {et~al.}(2012)\citenamefont {Xue},
  \citenamefont {Chen}, \citenamefont {Zhao}, \citenamefont {Guo},\ and\
  \citenamefont {Tang}}]{XCZ:IC12}%
  \BibitemOpen
  \bibfield  {author} {\bibinfo {author} {\bibfnamefont {S.}~\bibnamefont
  {Xue}}, \bibinfo {author} {\bibfnamefont {X.-H.}\ \bibnamefont {Chen}},
  \bibinfo {author} {\bibfnamefont {L.}~\bibnamefont {Zhao}}, \bibinfo {author}
  {\bibfnamefont {Y.-N.}\ \bibnamefont {Guo}}, \ and\ \bibinfo {author}
  {\bibfnamefont {J.}~\bibnamefont {Tang}},\ }\enquote {\bibinfo {title} {Two
  Bulky-Decorated Triangular Dysprosium Aggregates Conserving Vortex-Spin
  Structure},}\ \href {\doibase 10.1021/ic301785v} {\bibfield  {journal}
  {\bibinfo  {journal} {Inorg. Chem.}\ }\textbf {\bibinfo {volume} {51}},\
  \bibinfo {pages} {13264} (\bibinfo {year} {2012})}\BibitemShut {NoStop}%
\bibitem [{\citenamefont {Das}\ \emph {et~al.}(2015)\citenamefont {Das},
  \citenamefont {Vaidya}, \citenamefont {Gupta}, \citenamefont {Frost},
  \citenamefont {Righi}, \citenamefont {Brechin}, \citenamefont {Affronte},
  \citenamefont {Rajaraman},\ and\ \citenamefont {Shanmugam}}]{DVG:CAEJ15}%
  \BibitemOpen
  \bibfield  {author} {\bibinfo {author} {\bibfnamefont {C.}~\bibnamefont
  {Das}}, \bibinfo {author} {\bibfnamefont {S.}~\bibnamefont {Vaidya}},
  \bibinfo {author} {\bibfnamefont {T.}~\bibnamefont {Gupta}}, \bibinfo
  {author} {\bibfnamefont {J.~M.}\ \bibnamefont {Frost}}, \bibinfo {author}
  {\bibfnamefont {M.}~\bibnamefont {Righi}}, \bibinfo {author} {\bibfnamefont
  {E.~K.}\ \bibnamefont {Brechin}}, \bibinfo {author} {\bibfnamefont
  {M.}~\bibnamefont {Affronte}}, \bibinfo {author} {\bibfnamefont
  {G.}~\bibnamefont {Rajaraman}}, \ and\ \bibinfo {author} {\bibfnamefont
  {M.}~\bibnamefont {Shanmugam}},\ }\enquote {\bibinfo {title} {Single-Molecule
  Magnetism, Enhanced Magnetocaloric Effect, and Toroidal Magnetic Moments in a
  Family of Ln4 Squares},}\ \href {\doibase
  https://doi.org/10.1002/chem.201502720} {\bibfield  {journal} {\bibinfo
  {journal} {Chem. Eur. J.}\ }\textbf {\bibinfo {volume} {21}},\ \bibinfo
  {pages} {15639} (\bibinfo {year} {2015})}\BibitemShut {NoStop}%
\bibitem [{\citenamefont {Fernandez~Garcia}\ \emph {et~al.}(2018)\citenamefont
  {Fernandez~Garcia}, \citenamefont {Guettas}, \citenamefont {Montigaud},
  \citenamefont {Larini}, \citenamefont {Sessoli}, \citenamefont {Totti},
  \citenamefont {Cador}, \citenamefont {Pilet},\ and\ \citenamefont
  {Le~Guennic}}]{GGM:ACIE18}%
  \BibitemOpen
  \bibfield  {author} {\bibinfo {author} {\bibfnamefont {G.}~\bibnamefont
  {Fernandez~Garcia}}, \bibinfo {author} {\bibfnamefont {D.}~\bibnamefont
  {Guettas}}, \bibinfo {author} {\bibfnamefont {V.}~\bibnamefont {Montigaud}},
  \bibinfo {author} {\bibfnamefont {P.}~\bibnamefont {Larini}}, \bibinfo
  {author} {\bibfnamefont {R.}~\bibnamefont {Sessoli}}, \bibinfo {author}
  {\bibfnamefont {F.}~\bibnamefont {Totti}}, \bibinfo {author} {\bibfnamefont
  {O.}~\bibnamefont {Cador}}, \bibinfo {author} {\bibfnamefont
  {G.}~\bibnamefont {Pilet}}, \ and\ \bibinfo {author} {\bibfnamefont
  {B.}~\bibnamefont {Le~Guennic}},\ }\enquote {\bibinfo {title} {A {Dy$_4$}
  Cubane: A New Member in the Single-Molecule Toroics Family},}\ \href
  {\doibase https://doi.org/10.1002/anie.201810156} {\bibfield  {journal}
  {\bibinfo  {journal} {Angew. Chem. Int. Ed.}\ }\textbf {\bibinfo {volume}
  {57}},\ \bibinfo {pages} {17089} (\bibinfo {year} {2018})}\BibitemShut
  {NoStop}%
\bibitem [{\citenamefont {Vignesh}\ \emph {et~al.}(2018)\citenamefont
  {Vignesh}, \citenamefont {Langley}, \citenamefont {Swain}, \citenamefont
  {Moubaraki}, \citenamefont {Damjanović}, \citenamefont {Wernsdorfer},
  \citenamefont {Rajaraman},\ and\ \citenamefont {Murray}}]{VLS:ACIE18}%
  \BibitemOpen
  \bibfield  {author} {\bibinfo {author} {\bibfnamefont {K.~R.}\ \bibnamefont
  {Vignesh}}, \bibinfo {author} {\bibfnamefont {S.~K.}\ \bibnamefont
  {Langley}}, \bibinfo {author} {\bibfnamefont {A.}~\bibnamefont {Swain}},
  \bibinfo {author} {\bibfnamefont {B.}~\bibnamefont {Moubaraki}}, \bibinfo
  {author} {\bibfnamefont {M.}~\bibnamefont {Damjanović}}, \bibinfo {author}
  {\bibfnamefont {W.}~\bibnamefont {Wernsdorfer}}, \bibinfo {author}
  {\bibfnamefont {G.}~\bibnamefont {Rajaraman}}, \ and\ \bibinfo {author}
  {\bibfnamefont {K.~S.}\ \bibnamefont {Murray}},\ }\enquote {\bibinfo {title}
  {Slow Magnetic Relaxation and Single-Molecule Toroidal Behaviour in a Family
  of Heptanuclear {Cr$^{\text{III}}$Ln$^{\text{III}}_6$} ({Ln}={Tb}, {Ho},
  {Er}) Complexes},}\ \href {\doibase https://doi.org/10.1002/anie.201711844}
  {\bibfield  {journal} {\bibinfo  {journal} {Angew. Chem. Int. Ed.}\ }\textbf
  {\bibinfo {volume} {57}},\ \bibinfo {pages} {779} (\bibinfo {year}
  {2018})}\BibitemShut {NoStop}%
\bibitem [{\citenamefont {Crabtree}\ and\ \citenamefont
  {Soncini}(2018)}]{CrA:PRB18}%
  \BibitemOpen
  \bibfield  {author} {\bibinfo {author} {\bibfnamefont {J.~M.}\ \bibnamefont
  {Crabtree}}\ and\ \bibinfo {author} {\bibfnamefont {A.}~\bibnamefont
  {Soncini}},\ }\enquote {\bibinfo {title} {Toroidal quantum states in
  molecular spin-frustrated triangular nanomagnets with weak spin-orbit
  coupling: Applications to molecular spintronics},}\ \href {\doibase
  10.1103/PhysRevB.98.094417} {\bibfield  {journal} {\bibinfo  {journal} {Phys.
  Rev. B}\ }\textbf {\bibinfo {volume} {98}},\ \bibinfo {pages} {094417}
  (\bibinfo {year} {2018})}\BibitemShut {NoStop}%
\bibitem [{\citenamefont {Langley}\ \emph {et~al.}(2019)\citenamefont
  {Langley}, \citenamefont {Vignesh}, \citenamefont {Gupta}, \citenamefont
  {Gartshore}, \citenamefont {Rajaraman}, \citenamefont {Forsyth},\ and\
  \citenamefont {Murray}}]{LVG:DT19}%
  \BibitemOpen
  \bibfield  {author} {\bibinfo {author} {\bibfnamefont {S.~K.}\ \bibnamefont
  {Langley}}, \bibinfo {author} {\bibfnamefont {K.~R.}\ \bibnamefont
  {Vignesh}}, \bibinfo {author} {\bibfnamefont {T.}~\bibnamefont {Gupta}},
  \bibinfo {author} {\bibfnamefont {C.~J.}\ \bibnamefont {Gartshore}}, \bibinfo
  {author} {\bibfnamefont {G.}~\bibnamefont {Rajaraman}}, \bibinfo {author}
  {\bibfnamefont {C.~M.}\ \bibnamefont {Forsyth}}, \ and\ \bibinfo {author}
  {\bibfnamefont {K.~S.}\ \bibnamefont {Murray}},\ }\enquote {\bibinfo {title}
  {New examples of triangular terbium({III}) and holmium({III}) and hexagonal
  dysprosium({III}) single molecule toroics},}\ \href {\doibase
  10.1039/C9DT02419K} {\bibfield  {journal} {\bibinfo  {journal} {Dalton
  Trans.}\ }\textbf {\bibinfo {volume} {48}},\ \bibinfo {pages} {15657}
  (\bibinfo {year} {2019})}\BibitemShut {NoStop}%
\bibitem [{\citenamefont {Rao}\ \emph {et~al.}(2020)\citenamefont {Rao},
  \citenamefont {Ashtree},\ and\ \citenamefont {Soncini}}]{RAS:PB20}%
  \BibitemOpen
  \bibfield  {author} {\bibinfo {author} {\bibfnamefont {S.}~\bibnamefont
  {Rao}}, \bibinfo {author} {\bibfnamefont {J.}~\bibnamefont {Ashtree}}, \ and\
  \bibinfo {author} {\bibfnamefont {A.}~\bibnamefont {Soncini}},\ }\enquote
  {\bibinfo {title} {Toroidal moment in a family of spin-frustrated
  heterometallic triangular nanomagnets without spin-orbit coupling:
  Applications in a molecular spintronics device},}\ \href {\doibase
  https://doi.org/10.1016/j.physb.2020.412237} {\bibfield  {journal} {\bibinfo
  {journal} {Physica B}\ }\textbf {\bibinfo {volume} {592}},\ \bibinfo {pages}
  {412237} (\bibinfo {year} {2020})}\BibitemShut {NoStop}%
\bibitem [{\citenamefont {Pavlyukh}(2020)}]{Pav:PRB20}%
  \BibitemOpen
  \bibfield  {author} {\bibinfo {author} {\bibfnamefont {Y.}~\bibnamefont
  {Pavlyukh}},\ }\enquote {\bibinfo {title} {Toroidal spin states in molecular
  magnets},}\ \href {\doibase 10.1103/PhysRevB.101.144408} {\bibfield
  {journal} {\bibinfo  {journal} {Phys. Rev. B}\ }\textbf {\bibinfo {volume}
  {101}},\ \bibinfo {pages} {144408} (\bibinfo {year} {2020})}\BibitemShut
  {NoStop}%
\bibitem [{\citenamefont {Zhang}\ \emph {et~al.}(2020)\citenamefont {Zhang},
  \citenamefont {Zhai}, \citenamefont {Qin}, \citenamefont {Ungur},
  \citenamefont {Nojiri},\ and\ \citenamefont {Zheng}}]{ZZQ:M20}%
  \BibitemOpen
  \bibfield  {author} {\bibinfo {author} {\bibfnamefont {H.-L.}\ \bibnamefont
  {Zhang}}, \bibinfo {author} {\bibfnamefont {Y.-Q.}\ \bibnamefont {Zhai}},
  \bibinfo {author} {\bibfnamefont {L.}~\bibnamefont {Qin}}, \bibinfo {author}
  {\bibfnamefont {L.}~\bibnamefont {Ungur}}, \bibinfo {author} {\bibfnamefont
  {H.}~\bibnamefont {Nojiri}}, \ and\ \bibinfo {author} {\bibfnamefont {Y.-Z.}\
  \bibnamefont {Zheng}},\ }\enquote {\bibinfo {title} {Single-Molecule Toroic
  Design through Magnetic Exchange Coupling},}\ \href {\doibase
  https://doi.org/10.1016/j.matt.2020.02.021} {\bibfield  {journal} {\bibinfo
  {journal} {Matter}\ }\textbf {\bibinfo {volume} {2}},\ \bibinfo {pages} {1481
  } (\bibinfo {year} {2020})}\BibitemShut {NoStop}%
\bibitem [{\citenamefont {Ashtree}\ \emph {et~al.}(2021)\citenamefont
  {Ashtree}, \citenamefont {Borilovic}, \citenamefont {Vignesh}, \citenamefont
  {Swain}, \citenamefont {Hamilton}, \citenamefont {Whyatt}, \citenamefont
  {Benjamin}, \citenamefont {Phonsri}, \citenamefont {Forsyth}, \citenamefont
  {Wernsdorfer}, \citenamefont {Soncini}, \citenamefont {Rajaraman},
  \citenamefont {Langley},\ and\ \citenamefont {Murray}}]{ABV:EJIC21}%
  \BibitemOpen
  \bibfield  {author} {\bibinfo {author} {\bibfnamefont {J.~M.}\ \bibnamefont
  {Ashtree}}, \bibinfo {author} {\bibfnamefont {I.}~\bibnamefont {Borilovic}},
  \bibinfo {author} {\bibfnamefont {K.~R.}\ \bibnamefont {Vignesh}}, \bibinfo
  {author} {\bibfnamefont {A.}~\bibnamefont {Swain}}, \bibinfo {author}
  {\bibfnamefont {S.~H.}\ \bibnamefont {Hamilton}}, \bibinfo {author}
  {\bibfnamefont {Y.~L.}\ \bibnamefont {Whyatt}}, \bibinfo {author}
  {\bibfnamefont {S.~L.}\ \bibnamefont {Benjamin}}, \bibinfo {author}
  {\bibfnamefont {W.}~\bibnamefont {Phonsri}}, \bibinfo {author} {\bibfnamefont
  {C.~M.}\ \bibnamefont {Forsyth}}, \bibinfo {author} {\bibfnamefont
  {W.}~\bibnamefont {Wernsdorfer}}, \bibinfo {author} {\bibfnamefont
  {A.}~\bibnamefont {Soncini}}, \bibinfo {author} {\bibfnamefont
  {G.}~\bibnamefont {Rajaraman}}, \bibinfo {author} {\bibfnamefont {S.~K.}\
  \bibnamefont {Langley}}, \ and\ \bibinfo {author} {\bibfnamefont {K.~S.}\
  \bibnamefont {Murray}},\ }\enquote {\bibinfo {title} {Tuning the
  Ferrotoroidic Coupling and Magnetic Hysteresis in Double-Triangle Complexes
  \{{Dy$_3$M$^{\text{III}}$Dy$_3$}\} via the {M$^{\text{III}}$}-linker},}\
  \href {\doibase 10.1002/ejic.202001082} {\bibfield  {journal} {\bibinfo
  {journal} {Eur. J. Inorg. Chem.}\ }\textbf {\bibinfo {volume} {5}},\ \bibinfo
  {pages} {435} (\bibinfo {year} {2021})}\BibitemShut {NoStop}%
\bibitem [{\citenamefont {Troiani}\ \emph {et~al.}(2012)\citenamefont
  {Troiani}, \citenamefont {Stepanenko},\ and\ \citenamefont
  {Loss}}]{TSL:PRB12}%
  \BibitemOpen
  \bibfield  {author} {\bibinfo {author} {\bibfnamefont {F.}~\bibnamefont
  {Troiani}}, \bibinfo {author} {\bibfnamefont {D.}~\bibnamefont {Stepanenko}},
  \ and\ \bibinfo {author} {\bibfnamefont {D.}~\bibnamefont {Loss}},\ }\enquote
  {\bibinfo {title} {Hyperfine-induced decoherence in triangular spin-cluster
  qubits},}\ \href {\doibase 10.1103/PhysRevB.86.161409} {\bibfield  {journal}
  {\bibinfo  {journal} {Phys. Rev. B}\ }\textbf {\bibinfo {volume} {86}},\
  \bibinfo {pages} {161409} (\bibinfo {year} {2012})}\BibitemShut {NoStop}%
\bibitem [{\citenamefont {Garlatti}\ \emph {et~al.}(2021)\citenamefont
  {Garlatti}, \citenamefont {Chiesa}, \citenamefont {Bonfa}, \citenamefont
  {Macaluso}, \citenamefont {Onuorah}, \citenamefont {Parmar}, \citenamefont
  {Ding}, \citenamefont {Zheng}, \citenamefont {Giansiracusa}, \citenamefont
  {Reta}, \citenamefont {Pavarini}, \citenamefont {Guidi}, \citenamefont
  {Mills}, \citenamefont {Chilton}, \citenamefont {Winpenny}, \citenamefont
  {Santini},\ and\ \citenamefont {Carretta}}]{GCB:JPCL21}%
  \BibitemOpen
  \bibfield  {author} {\bibinfo {author} {\bibfnamefont {E.}~\bibnamefont
  {Garlatti}}, \bibinfo {author} {\bibfnamefont {A.}~\bibnamefont {Chiesa}},
  \bibinfo {author} {\bibfnamefont {P.}~\bibnamefont {Bonfa}}, \bibinfo
  {author} {\bibfnamefont {E.}~\bibnamefont {Macaluso}}, \bibinfo {author}
  {\bibfnamefont {I.~J.}\ \bibnamefont {Onuorah}}, \bibinfo {author}
  {\bibfnamefont {V.~S.}\ \bibnamefont {Parmar}}, \bibinfo {author}
  {\bibfnamefont {Y.-S.}\ \bibnamefont {Ding}}, \bibinfo {author}
  {\bibfnamefont {Y.-Z.}\ \bibnamefont {Zheng}}, \bibinfo {author}
  {\bibfnamefont {M.~J.}\ \bibnamefont {Giansiracusa}}, \bibinfo {author}
  {\bibfnamefont {D.}~\bibnamefont {Reta}}, \bibinfo {author} {\bibfnamefont
  {E.}~\bibnamefont {Pavarini}}, \bibinfo {author} {\bibfnamefont
  {T.}~\bibnamefont {Guidi}}, \bibinfo {author} {\bibfnamefont {D.~P.}\
  \bibnamefont {Mills}}, \bibinfo {author} {\bibfnamefont {N.~F.}\ \bibnamefont
  {Chilton}}, \bibinfo {author} {\bibfnamefont {R.~E.~P.}\ \bibnamefont
  {Winpenny}}, \bibinfo {author} {\bibfnamefont {P.}~\bibnamefont {Santini}}, \
  and\ \bibinfo {author} {\bibfnamefont {S.}~\bibnamefont {Carretta}},\
  }\enquote {\bibinfo {title} {A Cost-Effective Semi-Ab Initio Approach to
  Model Relaxation in Rare-Earth Single-Molecule Magnets},}\ \href {\doibase
  10.1021/acs.jpclett.1c02367} {\bibfield  {journal} {\bibinfo  {journal} {J.
  Phys. Chem. Lett.}\ }\textbf {\bibinfo {volume} {12}},\ \bibinfo {pages}
  {8826} (\bibinfo {year} {2021})}\BibitemShut {NoStop}%
\bibitem [{\citenamefont {Vorndamme}\ and\ \citenamefont
  {Schnack}(2020)}]{VoS:PRB20}%
  \BibitemOpen
  \bibfield  {author} {\bibinfo {author} {\bibfnamefont {P.}~\bibnamefont
  {Vorndamme}}\ and\ \bibinfo {author} {\bibfnamefont {J.}~\bibnamefont
  {Schnack}},\ }\enquote {\bibinfo {title} {Decoherence of a singlet-triplet
  superposition state under dipolar interactions of an uncorrelated
  environment},}\ \href {\doibase 10.1103/PhysRevB.101.075101} {\bibfield
  {journal} {\bibinfo  {journal} {Phys. Rev. B}\ }\textbf {\bibinfo {volume}
  {101}},\ \bibinfo {pages} {075101} (\bibinfo {year} {2020})}\BibitemShut
  {NoStop}%
\bibitem [{\citenamefont {Irl{\"a}nder}\ \emph {et~al.}(2021)\citenamefont
  {Irl{\"a}nder}, \citenamefont {Schmidt},\ and\ \citenamefont
  {Schnack}}]{ISS:EPJB21}%
  \BibitemOpen
  \bibfield  {author} {\bibinfo {author} {\bibfnamefont {K.}~\bibnamefont
  {Irl{\"a}nder}}, \bibinfo {author} {\bibfnamefont {H.-J.}\ \bibnamefont
  {Schmidt}}, \ and\ \bibinfo {author} {\bibfnamefont {J.}~\bibnamefont
  {Schnack}},\ }\enquote {\bibinfo {title} {Supersymmetric spin--phonon
  coupling prevents odd integer spins from quantum tunneling},}\ \href
  {\doibase 10.1140/epjb/s10051-021-00073-3} {\bibfield  {journal} {\bibinfo
  {journal} {Eur. Phys. J. B}\ }\textbf {\bibinfo {volume} {94}},\ \bibinfo
  {pages} {68} (\bibinfo {year} {2021})}\BibitemShut {NoStop}%
\bibitem [{Note1()}]{Note1}%
  \BibitemOpen
  \bibinfo {note} {${\protect \mathbf D}_i$ should not be confused with a
  similar symbol denoting the Dzyaloshinskii-Moriya vector. The later is a
  vector and connects two spins.}\BibitemShut {Stop}%
\bibitem [{\citenamefont {Vignesh}\ \emph {et~al.}(2017)\citenamefont
  {Vignesh}, \citenamefont {Soncini}, \citenamefont {Langley}, \citenamefont
  {Wernsdorfer}, \citenamefont {Murray},\ and\ \citenamefont
  {Rajaraman}}]{VSL:NC17}%
  \BibitemOpen
  \bibfield  {author} {\bibinfo {author} {\bibfnamefont {K.~R.}\ \bibnamefont
  {Vignesh}}, \bibinfo {author} {\bibfnamefont {A.}~\bibnamefont {Soncini}},
  \bibinfo {author} {\bibfnamefont {S.~K.}\ \bibnamefont {Langley}}, \bibinfo
  {author} {\bibfnamefont {W.}~\bibnamefont {Wernsdorfer}}, \bibinfo {author}
  {\bibfnamefont {K.~S.}\ \bibnamefont {Murray}}, \ and\ \bibinfo {author}
  {\bibfnamefont {G.}~\bibnamefont {Rajaraman}},\ }\enquote {\bibinfo {title}
  {Ferrotoroidic ground state in a heterometallic
  {{\{Cr$^{\text{III}}$Dy$^{\text{III}}_6$\}}} complex displaying slow magnetic
  relaxation},}\ \href {\doibase 10.1038/s41467-017-01102-5} {\bibfield
  {journal} {\bibinfo  {journal} {Nat. Commun.}\ }\textbf {\bibinfo {volume}
  {8}},\ \bibinfo {pages} {1023} (\bibinfo {year} {2017})}\BibitemShut
  {NoStop}%
\bibitem [{\citenamefont {Balcerzak}(2022)}]{Bal:JMMM22}%
  \BibitemOpen
  \bibfield  {author} {\bibinfo {author} {\bibfnamefont {T.}~\bibnamefont
  {Balcerzak}},\ }\enquote {\bibinfo {title} {Exact studies of the ground-state
  properties of the anisotropic Heisenberg dimer with spin S=1},}\ \href
  {\doibase https://doi.org/10.1016/j.jmmm.2022.169419} {\bibfield  {journal}
  {\bibinfo  {journal} {J. Magn. Magn. Mater.}\ }\textbf {\bibinfo {volume}
  {556}},\ \bibinfo {pages} {169419} (\bibinfo {year} {2022})}\BibitemShut
  {NoStop}%
\bibitem [{\citenamefont {Hoeke}\ \emph {et~al.}(2012)\citenamefont {Hoeke},
  \citenamefont {Heidemeier}, \citenamefont {Krickemeyer}, \citenamefont
  {Stammler}, \citenamefont {B{\"o}gge}, \citenamefont {Schnack}, \citenamefont
  {Postnikov},\ and\ \citenamefont {Glaser}}]{HHK:IC12}%
  \BibitemOpen
  \bibfield  {author} {\bibinfo {author} {\bibfnamefont {V.}~\bibnamefont
  {Hoeke}}, \bibinfo {author} {\bibfnamefont {M.}~\bibnamefont {Heidemeier}},
  \bibinfo {author} {\bibfnamefont {E.}~\bibnamefont {Krickemeyer}}, \bibinfo
  {author} {\bibfnamefont {A.}~\bibnamefont {Stammler}}, \bibinfo {author}
  {\bibfnamefont {H.}~\bibnamefont {B{\"o}gge}}, \bibinfo {author}
  {\bibfnamefont {J.}~\bibnamefont {Schnack}}, \bibinfo {author} {\bibfnamefont
  {A.}~\bibnamefont {Postnikov}}, \ and\ \bibinfo {author} {\bibfnamefont
  {T.}~\bibnamefont {Glaser}},\ }\enquote {\bibinfo {title} {Environmental
  Influence on the Single-Molecule Magnet Behavior of [MnIII6CrIII]3+:
  Molecular Symmetry versus Solid-State Effects},}\ \href
  {http://pubs.acs.org/doi/abs/10.1021/ic301406j} {\bibfield  {journal}
  {\bibinfo  {journal} {Inorg. Chem.}\ }\textbf {\bibinfo {volume} {51}},\
  \bibinfo {pages} {10929} (\bibinfo {year} {2012})}\BibitemShut {NoStop}%
\bibitem [{\citenamefont {Glaser}\ \emph {et~al.}(2015)\citenamefont {Glaser},
  \citenamefont {Hoeke}, \citenamefont {Gieb}, \citenamefont {Schnack},
  \citenamefont {Schr{\"o}der},\ and\ \citenamefont {M{\"u}ller}}]{GHG:CCR15}%
  \BibitemOpen
  \bibfield  {author} {\bibinfo {author} {\bibfnamefont {T.}~\bibnamefont
  {Glaser}}, \bibinfo {author} {\bibfnamefont {V.}~\bibnamefont {Hoeke}},
  \bibinfo {author} {\bibfnamefont {K.}~\bibnamefont {Gieb}}, \bibinfo {author}
  {\bibfnamefont {J.}~\bibnamefont {Schnack}}, \bibinfo {author} {\bibfnamefont
  {C.}~\bibnamefont {Schr{\"o}der}}, \ and\ \bibinfo {author} {\bibfnamefont
  {P.}~\bibnamefont {M{\"u}ller}},\ }\enquote {\bibinfo {title} {Quantum
  tunneling of the magnetization in {[Mn$^{\text{III}}_6$M]$^{3+}$ (M$=$
  Cr$^{\text{III}}$, Mn$^{\text{III}}$)} {SMM}s: Impact of molecular and
  crystal symmetry},}\ \href {\doibase
  http://dx.doi.org/10.1016/j.ccr.2014.12.001} {\bibfield  {journal} {\bibinfo
  {journal} {Coord. Chem. Rev.}\ }\textbf {\bibinfo {volume} {289-290}},\
  \bibinfo {pages} {261} (\bibinfo {year} {2015})}\BibitemShut {NoStop}%
\bibitem [{\citenamefont {Pister}(2022)}]{Pis:M22}%
  \BibitemOpen
  \bibfield  {author} {\bibinfo {author} {\bibfnamefont {D.}~\bibnamefont
  {Pister}},\ }\emph {\bibinfo {title} {Toroidale Momente in anisotropen
  Spinsystemen}},\ \href@noop {} {\bibinfo {type} {Master thesis}},\ \bibinfo
  {school} {Bielefeld University, Faculty of Physics} (\bibinfo {year}
  {2022})\BibitemShut {NoStop}%
\bibitem [{\citenamefont {Lippert}\ \emph {et~al.}(2017)\citenamefont
  {Lippert}, \citenamefont {Mukherjee}, \citenamefont {Broschinski},
  \citenamefont {Lippert}, \citenamefont {Walleck}, \citenamefont {Stammler},
  \citenamefont {B{\"o}gge}, \citenamefont {Schnack},\ and\ \citenamefont
  {Glaser}}]{LMB:IC17}%
  \BibitemOpen
  \bibfield  {author} {\bibinfo {author} {\bibfnamefont {K.-A.}\ \bibnamefont
  {Lippert}}, \bibinfo {author} {\bibfnamefont {C.}~\bibnamefont {Mukherjee}},
  \bibinfo {author} {\bibfnamefont {J.-P.}\ \bibnamefont {Broschinski}},
  \bibinfo {author} {\bibfnamefont {Y.}~\bibnamefont {Lippert}}, \bibinfo
  {author} {\bibfnamefont {S.}~\bibnamefont {Walleck}}, \bibinfo {author}
  {\bibfnamefont {A.}~\bibnamefont {Stammler}}, \bibinfo {author}
  {\bibfnamefont {H.}~\bibnamefont {B{\"o}gge}}, \bibinfo {author}
  {\bibfnamefont {J.}~\bibnamefont {Schnack}}, \ and\ \bibinfo {author}
  {\bibfnamefont {T.}~\bibnamefont {Glaser}},\ }\enquote {\bibinfo {title}
  {Suppression of Magnetic Quantum Tunneling in a Chiral Single-Molecule Magnet
  by Ferromagnetic Interactions},}\ \href {\doibase
  10.1021/acs.inorgchem.7b02453} {\bibfield  {journal} {\bibinfo  {journal}
  {Inorg. Chem.}\ }\textbf {\bibinfo {volume} {56}},\ \bibinfo {pages} {15119}
  (\bibinfo {year} {2017})}\BibitemShut {NoStop}%
\bibitem [{\citenamefont {Venne}\ \emph {et~al.}(2019)\citenamefont {Venne},
  \citenamefont {Feldscher}, \citenamefont {Walleck}, \citenamefont {Stammler},
  \citenamefont {B{\"o}gge}, \citenamefont {Schnack},\ and\ \citenamefont
  {Glaser}}]{BFW:CEJ19}%
  \BibitemOpen
  \bibfield  {author} {\bibinfo {author} {\bibfnamefont {J.-P.}\ \bibnamefont
  {Venne}}, \bibinfo {author} {\bibfnamefont {B.}~\bibnamefont {Feldscher}},
  \bibinfo {author} {\bibfnamefont {S.}~\bibnamefont {Walleck}}, \bibinfo
  {author} {\bibfnamefont {A.}~\bibnamefont {Stammler}}, \bibinfo {author}
  {\bibfnamefont {H.}~\bibnamefont {B{\"o}gge}}, \bibinfo {author}
  {\bibfnamefont {J.}~\bibnamefont {Schnack}}, \ and\ \bibinfo {author}
  {\bibfnamefont {T.}~\bibnamefont {Glaser}},\ }\enquote {\bibinfo {title}
  {Rational Improvement of Single-Molecule Magnets by Enforcing Ferromagnetic
  Interactions},}\ \href {\doibase 10.1002/chem.201805543} {\bibfield
  {journal} {\bibinfo  {journal} {Chem. Eur. J.}\ }\textbf {\bibinfo {volume}
  {25}},\ \bibinfo {pages} {4992} (\bibinfo {year} {2019})}\BibitemShut
  {NoStop}%
\bibitem [{\citenamefont {Oldengott}\ \emph {et~al.}(2020)\citenamefont
  {Oldengott}, \citenamefont {Schnack},\ and\ \citenamefont
  {Glaser}}]{OSG:EJIC20}%
  \BibitemOpen
  \bibfield  {author} {\bibinfo {author} {\bibfnamefont {J.~C.}\ \bibnamefont
  {Oldengott}}, \bibinfo {author} {\bibfnamefont {J.}~\bibnamefont {Schnack}},
  \ and\ \bibinfo {author} {\bibfnamefont {T.}~\bibnamefont {Glaser}},\
  }\enquote {\bibinfo {title} {Optimization of Single-Molecule Magnets by
  Suppression of Quantum Tunneling of the Magnetization},}\ \href {\doibase
  10.1002/ejic.202000507} {\bibfield  {journal} {\bibinfo  {journal} {Eur. J.
  Inorg. Chem.}\ }\textbf {\bibinfo {volume} {2020}},\ \bibinfo {pages} {3222}
  (\bibinfo {year} {2020})}\BibitemShut {NoStop}%
\bibitem [{\citenamefont {Hoeke}\ \emph {et~al.}(2014)\citenamefont {Hoeke},
  \citenamefont {Stammler}, \citenamefont {B{\"o}gge}, \citenamefont
  {Schnack},\ and\ \citenamefont {Glaser}}]{HSB:IC14}%
  \BibitemOpen
  \bibfield  {author} {\bibinfo {author} {\bibfnamefont {V.}~\bibnamefont
  {Hoeke}}, \bibinfo {author} {\bibfnamefont {A.}~\bibnamefont {Stammler}},
  \bibinfo {author} {\bibfnamefont {H.}~\bibnamefont {B{\"o}gge}}, \bibinfo
  {author} {\bibfnamefont {J.}~\bibnamefont {Schnack}}, \ and\ \bibinfo
  {author} {\bibfnamefont {T.}~\bibnamefont {Glaser}},\ }\enquote {\bibinfo
  {title} {Strong and Anisotropic Superexchange in the Single-Molecule Magnet
  (SMM) {[Mn$_6^{\text{III}}$Os$^{\text{III}}]^{3+}$}: Promoting {SMM} Behavior
  through 3d-5d Transition Metal Substitution},}\ \href
  {http://pubs.acs.org/doi/abs/10.1021/ic4022068} {\bibfield  {journal}
  {\bibinfo  {journal} {Inorg. Chem.}\ }\textbf {\bibinfo {volume} {53}},\
  \bibinfo {pages} {257} (\bibinfo {year} {2014})}\BibitemShut {NoStop}%
\bibitem [{\citenamefont {Kitaev}(2006)}]{Kit:AP06}%
  \BibitemOpen
  \bibfield  {author} {\bibinfo {author} {\bibfnamefont {A.}~\bibnamefont
  {Kitaev}},\ }\enquote {\bibinfo {title} {Anyons in an exactly solved model
  and beyond},}\ \href {\doibase http://dx.doi.org/10.1016/j.aop.2005.10.005}
  {\bibfield  {journal} {\bibinfo  {journal} {Ann. Phys.}\ }\textbf {\bibinfo
  {volume} {321}},\ \bibinfo {pages} {2 } (\bibinfo {year} {2006})}\BibitemShut
  {NoStop}%
\bibitem [{\citenamefont {Jackeli}\ and\ \citenamefont
  {Khaliullin}(2009)}]{JaK:PRL09}%
  \BibitemOpen
  \bibfield  {author} {\bibinfo {author} {\bibfnamefont {G.}~\bibnamefont
  {Jackeli}}\ and\ \bibinfo {author} {\bibfnamefont {G.}~\bibnamefont
  {Khaliullin}},\ }\enquote {\bibinfo {title} {Mott Insulators in the Strong
  Spin-Orbit Coupling Limit: From {Heisenberg} to a Quantum Compass and
  {Kitaev} Models},}\ \href {\doibase 10.1103/PhysRevLett.102.017205}
  {\bibfield  {journal} {\bibinfo  {journal} {Phys. Rev. Lett.}\ }\textbf
  {\bibinfo {volume} {102}},\ \bibinfo {pages} {017205} (\bibinfo {year}
  {2009})}\BibitemShut {NoStop}%
\bibitem [{\citenamefont {Winter}\ \emph {et~al.}(2016)\citenamefont {Winter},
  \citenamefont {Li}, \citenamefont {Jeschke},\ and\ \citenamefont
  {Valent\'{\i}}}]{WLJ:PRB16}%
  \BibitemOpen
  \bibfield  {author} {\bibinfo {author} {\bibfnamefont {S.~M.}\ \bibnamefont
  {Winter}}, \bibinfo {author} {\bibfnamefont {Y.}~\bibnamefont {Li}}, \bibinfo
  {author} {\bibfnamefont {H.~O.}\ \bibnamefont {Jeschke}}, \ and\ \bibinfo
  {author} {\bibfnamefont {R.}~\bibnamefont {Valent\'{\i}}},\ }\enquote
  {\bibinfo {title} {Challenges in design of {Kitaev} materials: Magnetic
  interactions from competing energy scales},}\ \href {\doibase
  10.1103/PhysRevB.93.214431} {\bibfield  {journal} {\bibinfo  {journal} {Phys.
  Rev. B}\ }\textbf {\bibinfo {volume} {93}},\ \bibinfo {pages} {214431}
  (\bibinfo {year} {2016})}\BibitemShut {NoStop}%
\bibitem [{\citenamefont {Zhang}\ \emph {et~al.}(2021)\citenamefont {Zhang},
  \citenamefont {Zhu}, \citenamefont {Go}, \citenamefont {Lux}, \citenamefont
  {dos Santos}, \citenamefont {Lounis}, \citenamefont {Su}, \citenamefont
  {Bl\"ugel},\ and\ \citenamefont {Mokrousov}}]{ZZG:PRB21}%
  \BibitemOpen
  \bibfield  {author} {\bibinfo {author} {\bibfnamefont {L.-C.}\ \bibnamefont
  {Zhang}}, \bibinfo {author} {\bibfnamefont {F.}~\bibnamefont {Zhu}}, \bibinfo
  {author} {\bibfnamefont {D.}~\bibnamefont {Go}}, \bibinfo {author}
  {\bibfnamefont {F.~R.}\ \bibnamefont {Lux}}, \bibinfo {author} {\bibfnamefont
  {F.~J.}\ \bibnamefont {dos Santos}}, \bibinfo {author} {\bibfnamefont
  {S.}~\bibnamefont {Lounis}}, \bibinfo {author} {\bibfnamefont
  {Y.}~\bibnamefont {Su}}, \bibinfo {author} {\bibfnamefont {S.}~\bibnamefont
  {Bl\"ugel}}, \ and\ \bibinfo {author} {\bibfnamefont {Y.}~\bibnamefont
  {Mokrousov}},\ }\enquote {\bibinfo {title} {Interplay of
  {Dzyaloshinskii-Moriya} and {Kitaev} interactions for magnonic properties of
  {Heisenberg-Kitaev} honeycomb ferromagnets},}\ \href {\doibase
  10.1103/PhysRevB.103.134414} {\bibfield  {journal} {\bibinfo  {journal}
  {Phys. Rev. B}\ }\textbf {\bibinfo {volume} {103}},\ \bibinfo {pages}
  {134414} (\bibinfo {year} {2021})}\BibitemShut {NoStop}%
\bibitem [{\citenamefont {Dzyaloshinsky}(1958)}]{Dzy:JPCS58}%
  \BibitemOpen
  \bibfield  {author} {\bibinfo {author} {\bibfnamefont {I.}~\bibnamefont
  {Dzyaloshinsky}},\ }\enquote {\bibinfo {title} {A thermodynamic theory of
  weak ferromagnetism of antiferromagnetics},}\ \href
  {http://dx.doi.org/10.1016/0022-3697(58)90076-3} {\bibfield  {journal}
  {\bibinfo  {journal} {J. Phys. Chem. Solids}\ }\textbf {\bibinfo {volume}
  {4}},\ \bibinfo {pages} {241} (\bibinfo {year} {1958})}\BibitemShut {NoStop}%
\bibitem [{\citenamefont {Moriya}(1960{\natexlab{a}})}]{Mor:PRL60}%
  \BibitemOpen
  \bibfield  {author} {\bibinfo {author} {\bibfnamefont {T.}~\bibnamefont
  {Moriya}},\ }\enquote {\bibinfo {title} {New mechanism of anisotropic
  superexchange interaction},}\ \href
  {http://dx.doi.org/10.1103/PhysRevLett.4.228} {\bibfield  {journal} {\bibinfo
   {journal} {Phys. Rev. Lett.}\ }\textbf {\bibinfo {volume} {4}},\ \bibinfo
  {pages} {228} (\bibinfo {year} {1960}{\natexlab{a}})}\BibitemShut {NoStop}%
\bibitem [{\citenamefont {Moriya}(1960{\natexlab{b}})}]{Mor:PR60}%
  \BibitemOpen
  \bibfield  {author} {\bibinfo {author} {\bibfnamefont {T.}~\bibnamefont
  {Moriya}},\ }\enquote {\bibinfo {title} {Anisotropic superexchange
  interaction and weak ferromagnetism},}\ \href
  {http://dx.doi.org/10.1103/PhysRev.120.91} {\bibfield  {journal} {\bibinfo
  {journal} {Phys. Rev.}\ }\textbf {\bibinfo {volume} {120}},\ \bibinfo {pages}
  {91} (\bibinfo {year} {1960}{\natexlab{b}})}\BibitemShut {NoStop}%
\bibitem [{\citenamefont {Grytsiuk}\ \emph {et~al.}(2020)\citenamefont
  {Grytsiuk}, \citenamefont {Hanke}, \citenamefont {Hoffmann}, \citenamefont
  {Bouaziz}, \citenamefont {Gomonay}, \citenamefont {Bihlmayer}, \citenamefont
  {Lounis}, \citenamefont {Mokrousov},\ and\ \citenamefont
  {Bl{\"u}gel}}]{GHH:NC20}%
  \BibitemOpen
  \bibfield  {author} {\bibinfo {author} {\bibfnamefont {S.}~\bibnamefont
  {Grytsiuk}}, \bibinfo {author} {\bibfnamefont {J.~P.}\ \bibnamefont {Hanke}},
  \bibinfo {author} {\bibfnamefont {M.}~\bibnamefont {Hoffmann}}, \bibinfo
  {author} {\bibfnamefont {J.}~\bibnamefont {Bouaziz}}, \bibinfo {author}
  {\bibfnamefont {O.}~\bibnamefont {Gomonay}}, \bibinfo {author} {\bibfnamefont
  {G.}~\bibnamefont {Bihlmayer}}, \bibinfo {author} {\bibfnamefont
  {S.}~\bibnamefont {Lounis}}, \bibinfo {author} {\bibfnamefont
  {Y.}~\bibnamefont {Mokrousov}}, \ and\ \bibinfo {author} {\bibfnamefont
  {S.}~\bibnamefont {Bl{\"u}gel}},\ }\enquote {\bibinfo {title}
  {Topological--chiral magnetic interactions driven by emergent orbital
  magnetism},}\ \href {\doibase 10.1038/s41467-019-14030-3} {\bibfield
  {journal} {\bibinfo  {journal} {Nat. Commun.}\ }\textbf {\bibinfo {volume}
  {11}},\ \bibinfo {pages} {511} (\bibinfo {year} {2020})}\BibitemShut
  {NoStop}%
\bibitem [{\citenamefont {Hoffmann}\ and\ \citenamefont
  {Bl\"ugel}(2020)}]{HoB:PRB20}%
  \BibitemOpen
  \bibfield  {author} {\bibinfo {author} {\bibfnamefont {M.}~\bibnamefont
  {Hoffmann}}\ and\ \bibinfo {author} {\bibfnamefont {S.}~\bibnamefont
  {Bl\"ugel}},\ }\enquote {\bibinfo {title} {Systematic derivation of realistic
  spin models for beyond-Heisenberg solids},}\ \href {\doibase
  10.1103/PhysRevB.101.024418} {\bibfield  {journal} {\bibinfo  {journal}
  {Phys. Rev. B}\ }\textbf {\bibinfo {volume} {101}},\ \bibinfo {pages}
  {024418} (\bibinfo {year} {2020})}\BibitemShut {NoStop}%
\bibitem [{\citenamefont {dos Santos~Dias}\ \emph {et~al.}(2021)\citenamefont
  {dos Santos~Dias}, \citenamefont {Brinker}, \citenamefont {L\'aszl\'offy},
  \citenamefont {Ny\'ari}, \citenamefont {Bl\"ugel}, \citenamefont {Szunyogh},\
  and\ \citenamefont {Lounis}}]{SBL:PRB21}%
  \BibitemOpen
  \bibfield  {author} {\bibinfo {author} {\bibfnamefont {M.}~\bibnamefont {dos
  Santos~Dias}}, \bibinfo {author} {\bibfnamefont {S.}~\bibnamefont {Brinker}},
  \bibinfo {author} {\bibfnamefont {A.}~\bibnamefont {L\'aszl\'offy}}, \bibinfo
  {author} {\bibfnamefont {B.}~\bibnamefont {Ny\'ari}}, \bibinfo {author}
  {\bibfnamefont {S.}~\bibnamefont {Bl\"ugel}}, \bibinfo {author}
  {\bibfnamefont {L.}~\bibnamefont {Szunyogh}}, \ and\ \bibinfo {author}
  {\bibfnamefont {S.}~\bibnamefont {Lounis}},\ }\enquote {\bibinfo {title}
  {Proper and improper chiral magnetic interactions},}\ \href {\doibase
  10.1103/PhysRevB.103.L140408} {\bibfield  {journal} {\bibinfo  {journal}
  {Phys. Rev. B}\ }\textbf {\bibinfo {volume} {103}},\ \bibinfo {pages}
  {L140408} (\bibinfo {year} {2021})}\BibitemShut {NoStop}%
\bibitem [{\citenamefont {Leavens}\ and\ \citenamefont
  {Aers}(1988)}]{LeA:PRB88}%
  \BibitemOpen
  \bibfield  {author} {\bibinfo {author} {\bibfnamefont {C.~R.}\ \bibnamefont
  {Leavens}}\ and\ \bibinfo {author} {\bibfnamefont {G.~C.}\ \bibnamefont
  {Aers}},\ }\enquote {\bibinfo {title} {Tunneling current density within
  {Tersoff} and {Hamann's} theory of the scanning tunneling microscope},}\
  \href {\doibase 10.1103/PhysRevB.38.7357} {\bibfield  {journal} {\bibinfo
  {journal} {Phys. Rev. B}\ }\textbf {\bibinfo {volume} {38}},\ \bibinfo
  {pages} {7357} (\bibinfo {year} {1988})}\BibitemShut {NoStop}%
\bibitem [{\citenamefont {Bae}\ \emph {et~al.}(2018)\citenamefont {Bae},
  \citenamefont {Yang}, \citenamefont {Willke}, \citenamefont {Choi},
  \citenamefont {Heinrich},\ and\ \citenamefont {Lutz}}]{BYW:SA18}%
  \BibitemOpen
  \bibfield  {author} {\bibinfo {author} {\bibfnamefont {Y.}~\bibnamefont
  {Bae}}, \bibinfo {author} {\bibfnamefont {K.}~\bibnamefont {Yang}}, \bibinfo
  {author} {\bibfnamefont {P.}~\bibnamefont {Willke}}, \bibinfo {author}
  {\bibfnamefont {T.}~\bibnamefont {Choi}}, \bibinfo {author} {\bibfnamefont
  {A.~J.}\ \bibnamefont {Heinrich}}, \ and\ \bibinfo {author} {\bibfnamefont
  {C.~P.}\ \bibnamefont {Lutz}},\ }\enquote {\bibinfo {title} {Enhanced quantum
  coherence in exchange coupled spins via singlet-triplet transitions},}\ \href
  {\doibase 10.1126/sciadv.aau4159} {\bibfield  {journal} {\bibinfo  {journal}
  {Science Advances}\ }\textbf {\bibinfo {volume} {4}},\ \bibinfo {pages}
  {eaau4159} (\bibinfo {year} {2018})}\BibitemShut {NoStop}%
\end{thebibliography}

%merlin.mbs apsrev4-1.bst; modified by Andreas Honecker for title output
%Control: key (0)
%Control: author (72) initials jnrlst
%Control: editor formatted (1) identically to author
%Control: production of article title (-1) disabled
%Control: page (0) single
%Control: year (1) truncated
%Control: production of eprint (0) enabled
%

\end{document}